\documentclass[aps,prb,twocolumn,groupedaddress,showpacs,superscriptaddress,amssymb,amsmath,noeprint]{revtex4-2}
\usepackage{graphicx}
\usepackage{dcolumn}
\usepackage{bm}
\usepackage{hyperref}
\usepackage{cleveref}
\usepackage{multirow}
\hypersetup{
    colorlinks=true,
    linkcolor=blue,
    urlcolor=blue,
    citecolor=blue
}

\newcommand{\la}{\langle}
\newcommand{\ra}{\rangle}

\newcommand{\tr}{\mathrm{Tr}}

\begin{document}

\title{Quantum trajectories and Page-curve entanglement dynamics}

\author{Katha Ganguly}
\email{katha.ganguly@students.iiserpune.ac.in}
\affiliation{Department of Physics, Indian Institute of Science Education and Research Pune, Dr. Homi Bhabha Road, Ward No. 8, NCL Colony, Pashan, Pune, Maharashtra 411008, India}

\author{Preethi Gopalakrishnan}
\email{preethi.g@icts.res.in } 
\affiliation{International Centre for Theoretical Sciences, Tata Institute of Fundamental Research,
Bangalore 560089, India}

\author{Atharva Naik}
\email{atharva.naik@icts.res.in} 
\affiliation{International Centre for Theoretical Sciences, Tata Institute of Fundamental Research,
Bangalore 560089, India}

\author{Bijay Kumar Agarwalla}
\email{bijay@iiserpune.ac.in}
\affiliation{Department of Physics, Indian Institute of Science Education and Research Pune, Dr. Homi Bhabha Road, Ward No. 8, NCL Colony, Pashan, Pune, Maharashtra 411008, India}

\author{Manas Kulkarni}
\email{manas.kulkarni@icts.res.in} 
\affiliation{International Centre for Theoretical Sciences, Tata Institute of Fundamental Research,
Bangalore 560089, India}
\thanks{\textsuperscript{†}These authors contributed equally to this work.}

\date{\today} 

\begin{abstract}
We consider time dynamics of entanglement entropy
between a filled fermionic system and an empty reservoir. We consider scenarios (i) where the system is subjected to a dephasing mechanism and the reservoir is clean, thereby emulating expansion of effectively interacting fermions in vacuum, and (ii) where both the system and the reservoir are subjected to dephasing and thereby enabling us to address how the entanglement between the part of the effectively interacting system and its complement evolves in time. We consider two different kinds of quantum trajectory approaches, namely stochastic unitary unraveling and quantum state diffusion. For both protocols, we observe and characterize the full Page curve-like dynamics for the entanglement entropy.  Depending on the protocol and the setup, we observe very distinct characteristics of the Page curve and the associated Page time and Page value. We also compute the number of fermions leaking to the reservoir and the associated current and shed light on their plausible connections with entanglement entropy. Our findings are expected to hold for a wide variety of generic interacting quantum systems. 
\end{abstract}

\maketitle

\textit{Introduction.}-- Understanding quantum entanglement~\cite{Horodecki2009,nielsen2010quantum,Abanin2019} through its various measures, such as von Neumann entanglement entropy~\cite{Plenio2010}, negativity~\cite{PhysRevLett.109.130502,PRXQuantum.2.030313,negativity_schiro} has gained significant interest in several branches of physics, ranging from condensed matter~\cite{GioevKilch2006,KitaevPreskill,Swingle2010,Jiang2012,rachel2015quantum,LAFLORENCIE20161} to high-energy physics~\cite{Nishioka_2009,Calabrese_2009,RevModPhys.90.035007}. It brings together disparate fields such as quantum information~\cite{preskill1998lecture}, many-body~\cite{Rosario_entanglement2008,PhysRevLett.109.020504,Srivastava_2024}, and black hole physics~\cite{RT2006,blackhole1}.  In addition to enormous theoretical developments, there have been several advances in experiments~\cite{rajibul2015,Vedika2019,PeterZoller2019,Noel2022,experiment2,Nature2024} to understand temporal growth, system size scaling, and using entanglement as a resource for quantum information processing for many-body quantum systems. Moreover, the consequence of entanglement when unitary dynamics is interspersed with quantum measurement is an area that is actively explored. In fact, it is known to lead to phase transitions that are measurement-induced~\cite{PhysRevB.98.205136,10.21468/SciPostPhys.7.2.024,Fisher2019,AdamNahum2019,Sarang2020,PhysRevB.102.054302,PhysRevB.102.035119,PhysRevResearch.2.043072,PhysRevA.102.033316,PhysRevResearch.2.013022,PhysRevB.103.174303,PhysRevB.103.224210,sebastian2021,Norman2022,Sebastian2022,Keiji_saito_PRL_long-range,negativity_schiro}. 

One important aspect is exploring the full dynamics of entanglement, i.e., its initial growth, attainment of a maximum value, and subsequent decay. This is generally referred to as the Page curve, which is commonly discussed in the context of black-hole evaporation~\cite{Page1993,Page_2013,Piroli2020}. The entanglement, in this case, is between the black-hole and the radiation. Initially, this entanglement is zero. As the black-hole radiates, the entanglement between the black-hole and the radiation keeps increasing and at a certain Page time, this entanglement is maximum. After this time, the entanglement decreases and eventually is expected to go to zero when the black-hole fully evaporates. 
It is worth noting that this non-monotonic behaviour was evasive in Hawking’s semiclassical calculation~\cite{hawking75}, which sparked discussions on the information paradox~\cite{mathur2009information,maes2015no,Bertini_2018,almheiri2021entropy,raju2022lessons}. 
Very recently, analytically available platforms have been studied that mimic such dynamics of entanglement entropy~\cite{Kehrein2024, saha2024,Glatthard2024,glatthard2025}. These recent works are notably different from (i) several other works that studied nonequilibrium dynamics of entanglement but were not designed to capture the full Page curve dynamics~\cite{Calabrese_2005,Calabrese_2007,KH2013,HA2017,Calabrese2017,Scopa_2021}. This is because typically a bipartition system is considered where the ratio between the part of the system and its complement is non-negligible and (ii) several works which studied steady state entanglement as a function of subsystem fraction which also gives rise to Page curve-like behaviour~\cite{PhysRevLett.125.180604,Bhattacharjee2021,Mario2021,PRXQuantum.3.030201,PhysRevResearch.5.013044}.


A challenging question that is far less understood is the entanglement dynamics of the full Page curve in quantum systems, which are either inherently interacting or systems that are effectively interacting, such as systems subject to dephasing mechanisms. While inbuilt many-body quantum interactions become computationally formidable, a route to circumvent this challenge is to use dephasing probes that effectively mimic scattering effects caused by interactions~\cite{Znidaric_2010, liang2024dephasing,Longhi2024,Landi_review}. How this dephasing mechanism emulates interactions in a non-interacting system is well studied in the literature in the context of wave packet spreading as well as system size scaling of steady state current~\cite{Pastawski1990,Abhishek2007,Goold2021}. The typical non-interacting system shows the ballistic spreading of wave packet and ballistic system size scaling of the nonequilibrium steady state current~\cite{Landi_review}. The presence of a dephasing mechanism in non-interacting systems leads to diffusive scaling of wave packet dynamics~\cite{CFRoos2019,Ghosh_2024,PhysRevB.110.L081403} and steady state current~\cite{10.1063/1.4926395,Schiro_dephasing,Madhumita_transport}. This is a typical feature of chaotic quantum systems that includes many-body interactions. The necessary setup to observe all the essential features of the Page entanglement curve consists of a partition comprising of a finite system and an infinite environment. This makes this problem challenging as it involves a large Hilbert space by design.  

\begin{figure}
    \centering
    \includegraphics[width=8.8cm,height=6.3cm]{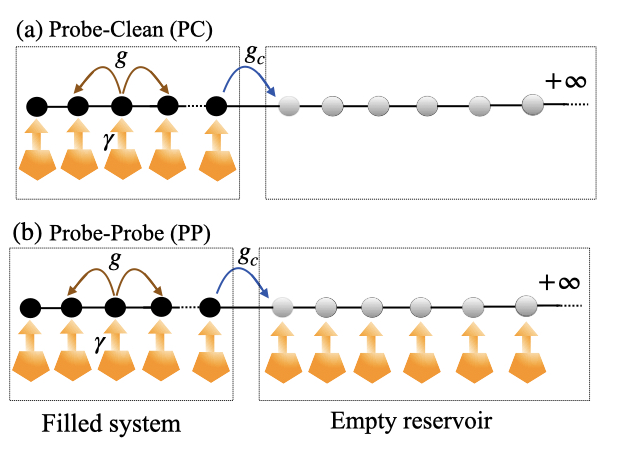}
    \caption{Schematic showing a finite-size filled fermionic system of length $L_S$ coupled with an infinite empty fermionic reservoir. We call the setup (a) as Probe-Clean (PC) setup, where only the system sites are subjected to dephasing probes with strength $\gamma$ but the reservoirs are not, and (b) as Probe-Probe (PP) setup, where the dephasing probes are present throughout the system and the reservoir lattice sites. The hopping strength for both the system and the reservoir is $g$, except at the junction of the system and the reservoir, where it is $g_c$. The orange shapes in both the figures generically denote dephasing probes. In Supplementary material~\cite{supp}, we provide details of two kinds of stochastic protocols that mimic the action of dephasing probes on an average.}
    \label{fig:Setup_schematic}
\end{figure}

In this work, we study Page curve-like dynamics of Entanglement Entropy (EE) in a Hamiltonian reservoir setup under different kinds of geometrical arrangement of probes depicted in schematic Fig.~\ref{fig:Setup_schematic}. For each of these geometry, we have considered two types of unraveling protocol. The central results are summarized in Table.~\ref{table_1}. We also provide compelling numerical evidence to support deep connections between EE and particle current between the system and the reservoir.

\textit{Setup.}--
The total Hamiltonian comprising of the system and the reservoir (see Fig.~\ref{fig:Setup_schematic}) is given by
\begin{equation}
\label{eq:h}
    \hat{H} = \sum_{i,j =1}^{L} h_{ij} \hat{c}^\dagger_{i} \hat{c}_{j},
\end{equation}
where $\hat{c}_i$ and $\hat{c}_i^\dagger$ are the fermionic annihilation and creation operator, respectively at site $i$ of the 1D chain. $h$ is the single-particle Hamiltonian which is a $L\times L$ matrix with matrix elements 
\begin{equation}
\label{eq:hel}
  h_{i,j} = -g\,(\delta_{i,j+1} + \delta_{i+1,j}), \,\,\, \forall \, i, j \neq L_{S}+1, L_S.   
\end{equation}

Here, $g$ is the nearest neighbor hopping strength within both the system and the reservoir. However, at the junction of the system and the reservoir, we assign a different hopping strength $g_c$. As manifested in Fig.~\ref{fig:Setup_schematic}, the setup is further subject to dephashing mechanisms. 
\begin{table}[]
\centering
\begin{tabular}{|c|c|c|c|}
\hline
Setup &Protocol & $t < t_p$ & $t > t_p$ \\ \cline{1-4}
 \multirow{2}{*}{PC  [Fig.~\ref{fig:Setup_schematic}a]} & SUU & $\sqrt{t}$ [Fig.~\ref{fig: PC}a] & $\ln(1/t) $[Fig.~\ref{fig: PC}a] \\ \cline{2-4}
                         & QSD & $\ln(t)$  [Fig.~\ref{fig: PC}b]& $\ln(1/t)$  [Fig.~\ref{fig: PC}b]\\ \hline
\multirow{2}{*}{PP  [Fig.~\ref{fig:Setup_schematic}b]} & SUU & $\sqrt{t}$  [Fig.~\ref{fig:PP}a]& $1/t^{0.4}$  [Fig.~\ref{fig:PP}a]\\ \cline{2-4}
                     & QSD & $\ln(\ln(t))$ [Fig.~\ref{fig:PP}b] & $1/\sqrt{t}$  [Fig.~\ref{fig:PP}b] \\ \hline
\end{tabular}
\caption{Table summarizing Entanglement Entropy (EE) $S(t)$ scaling under Stochastic Unitary Unraveling (SUU) and Quantum State Diffusion (QSD) protocols for the Probe-Clean (PC) [Fig.~\ref{fig:Setup_schematic}a] and the Probe-Probe (PP)[Fig.~\ref{fig:Setup_schematic}b] setup.}
\label{table_1}
\end{table}
The density matrix of the full system and reservoir takes the following Lindblad form,  
\begin{align}
    \frac{d\rho}{dt} = -i[\hat{H}, \rho] + \gamma\sum_{i=1}^{L_P} \left( \hat{n}_i \rho\, \hat{n}_i - \frac{1}{2}\{\hat{n}_i, \rho\} \right),
\label{eq:lindblad}
\end{align}
where $\hat{H}$ is the total Hamiltonian of the setup, given in Eq.~\eqref{eq:hel}, that includes the system (with size $L_S$) and the reservoir (with size $L_R$) with total number of sites $L=L_S+L_R$. $L_P$ is the total number of dephasing probes which is $L_S$ for the Probe-Clean (PC) case and $L$ for the Probe-Probe (PP) case and $\gamma$ is the dephasing strength.  
The Lindblad Master equation, given in Eq.~\eqref{eq:lindblad}, is well suited to study both the dynamics and steady states of the expectation values of observables that depend linearly on $\rho(t)$ i.e., $\la \hat{\mathcal{O}} \ra_t =\tr{[\rho(t) \, \hat{\mathcal{O}]}}$. However, quantities that are inherently nonlinear in $\rho(t)$ fall outside the realm of the Lindblad equation, and careful unraveling of the underlying density matrix evolution is warranted. In this regard, we consider two kinds of unraveling, namely, Stochastic Unitary Unraveling (SUU) and Quantum State Diffusion (QSD), the details of which are presented in the Supplementary material~\cite{supp}. Former is the onsite fluctuating noise added to the Hamiltonian thereby making the evolution stochastic while maintaining its unitarity~\cite{PhysRevLett.118.140403,PhysRevB.102.100301,PhysRevB.110.L081403,Giamarchi_transport2024}. Latter is the unraveling that mimics continuous weak measurement of the local particle number $\hat{n}_i$ at each site~\cite{sebastian2021,Giamarchi_transport2024}. While both these unraveling schemes upon suitable averaging yield the same Lindblad equation [Eq.~\eqref{eq:lindblad}], they can give rise to completely different physics for the quantities that depend nonlinearly on the state $\rho$. One such quantity is the Entanglement Entropy (EE), which is the primary focus of this work. We are interested in investigating the time dynamics of EE between the system and the reservoir in the PC and PP cases, shown in Fig.~\ref{fig:Setup_schematic}a and~\ref{fig:Setup_schematic}b, respectively. As the dephasing probe is well known for mimicking interaction, we study the PC setup to understand the entanglement dynamics in the expansion of interacting gas into a free vacuum. The PP setup is of interest because it addresses how a part of an interacting system is entangled with a large interacting reservoir. 

The procedure we adopt to calculate EE is the following: we first calculate the $L\times L_S$ dimensional matrix $U$, which is an isometry ($U^{\dagger}U=\mathbb{I}$). This is done for both the PC and PP cases under both kinds of unraveling, namely the SUU and the QSD~\cite{supp}. 
We then connect the correlation matrix with the matrix $U$ for each quantum trajectory via the relation,
\begin{align}
    C^{\xi}_{ij}(t)= \langle \hat{c}_i^{\dagger}(t) \hat{c}_j(t) \rangle = [U(t)\,U^{\dagger}(t)]_{ji}, \label{corr_mat}
\end{align}
where the superscript $\xi$ represents each quantum trajectory and the average is over the initial state. Since each quantum trajectory follows a Gaussian evolution, one can compute the evolution of EE from the correlation matrix as ~\cite{Peschel_2009,sharma,saha2024, supp,Rajdeep2023},
\begin{align}
    S_{\xi}(t)=-\sum_{k=1}^{L_S}\Big[\lambda_{k}\log_{2}\lambda_k+(1-\lambda_k)\log_2(1-\lambda_k)\Big], \label{EE}
\end{align}
where $\lambda_k \geq 0$ are the eigenvalues of the part of the correlation matrix that belongs to the system. Finally, we average over all the noise realizations/quantum trajectories to get $S(t)=\overline{S}_{\xi}(t)$. Additionally, from the correlation matrix one can extract the number of fermions leaking to the reservoir i.e., $\mathcal{N}_R(t)=\sum_{i=L_S+1}^{\infty}\overline{C_{ii}^{\xi}(t)}$ and the particle current $\mathcal{I}(t)=d\mathcal{N}_R/dt$. Note that quantities such as $\mathcal{N}_R(t)$ and $\mathcal{I}(t)$ can also be computed directly following the Lindblad equation in Eq.~\eqref{eq:lindblad}.

\begin{figure}
    \centering
    \includegraphics[width=0.49\columnwidth]{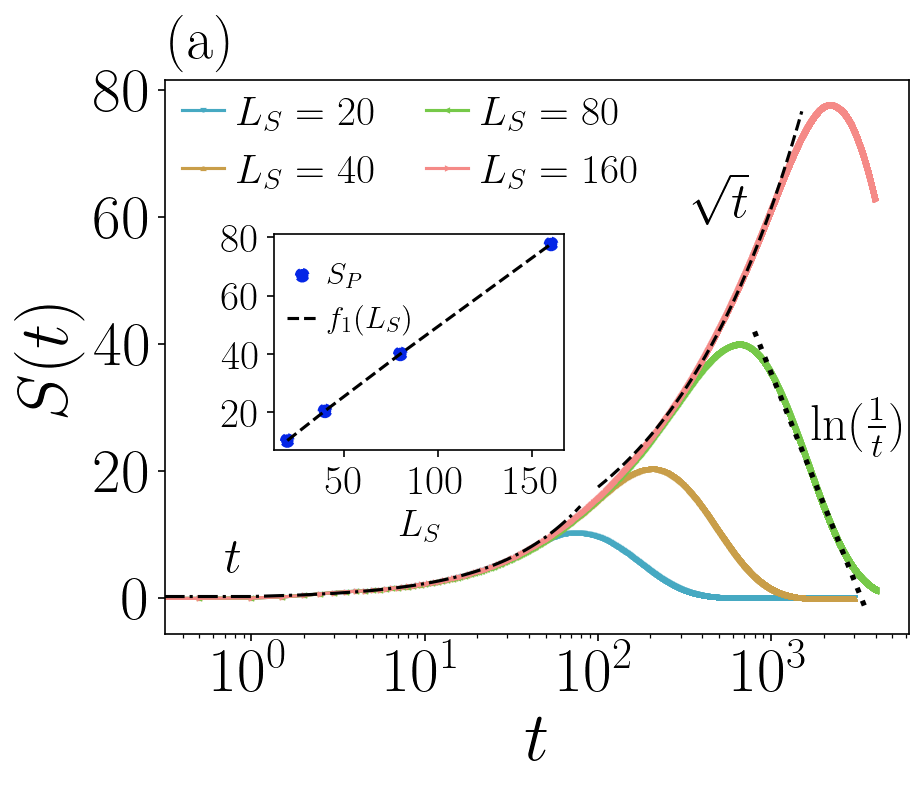}
    \includegraphics[width=0.49\columnwidth]{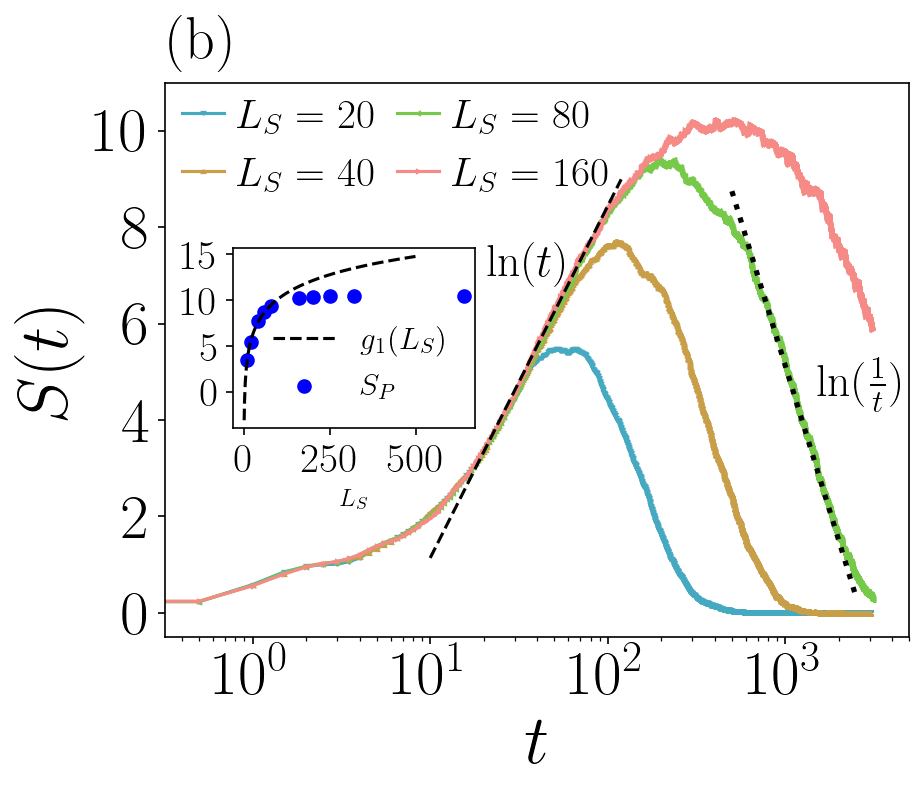}
    \caption{Plot of the entanglement entropy $S(t)$ with time $t$ for the Probe-Clean (PC) setup for the parameters $\gamma =  0.1$, $g=0.5$, $g_c  = 0.4$. The data is averaged over 100 trajectories. (a) SUU case: the entanglement entropy shows a crossover from initial linear growth in time to a diffusive $\sqrt{t}$ growth in the increasing part whereas in the decreasing part, it shows a logarithmic fall, (b) QSD case: the entanglement entropy shows a $\ln t$ growth in the increasing part and a logarithmic fall in the decreasing part. The inset shows the Page value $S_P$ vs system size $L_{S}$ plot which shows volume law scaling in (a) with $f_{1}(L_S)=aL_S+b$ where $a=0.5$ and $b=1.1$ and sub-volume law in (b) with $g_1(L_S) = a\ln{(L_S)}+b$ where $a  = 6.2$ and $b = -2.3$. Unlike in SUU, for the QSD case [(b)], a clear crossover from sub-volume to area law $(L_S^{0})$ scaling is observed in the Page value.}
    \label{fig: PC}
\end{figure}

We first discuss the temporal growth and subsequent decay of EE (which we refer to as the Page curve) for different cases, starting with the domain wall initial state, i.e., the system is completely filled and the bath is empty. Such a state is Gaussian. The system and reservoir sizes are chosen so as to ensure that the entire Page curve is realized and the boundary effect due to the finite size of the reservoir has not kicked in. It should be noted that as long as the reservoir is large enough to prevent finite-size effects, our results are independent of the reservoir size. We summarize the main findings in Table.~\ref{table_1}. A typical time evolution of EE is of the following form: There is initially a temporal growth region of EE, starting with zero value, which finally maximizes to a Page value $S_P$ at a Page time $t_P$. The EE then decreases, eventually going all the way to zero. This dynamics of entanglement is concomitant with fermions exiting the system into the reservoir and eventually resulting in an empty system. For all our numerical simulations for EE, we use $\gamma=0.1$ and average over $100$ noise trajectories. We have confirmed the convergence of our results with respect to increasing number of trajectories. The maximum reservoir size that we consider for our simulations is $L_R =6000$. In Fig.~\ref{fig: PC}, we discuss the PC case shown schematically in Fig.~\ref{fig:Setup_schematic}a both for SUU and QSD protocol. In the SUU case, i.e., Fig.~\ref{fig: PC}a, EE is characterized first by a very early time linear growth after which the dephasing mechanism kicks in around $t\sim O(1/\gamma)$. We then observe a clear crossover to diffusive $\sqrt{t}$ growth for EE. This $\sqrt{t}$ behaviour is somewhat intuitive given the fact that the number of particles $\mathcal{N}_R(t)$ exiting the system also scales in the same manner. It is these fermions in the reservoir that get entangled with those in the system. This relation between the temporal growth of EE and $\mathcal{N}_R$ is by no means obvious nor necessary. This is because the number of particles that exit the system and the amount of entanglement they develop with those remaining in the system may not have a straightforward connection, as we will elaborate later. The entanglement growth will eventually peak at Page value $S_P$ at the Page time $t_P$. Subsequently, EE decays all the way to zero as $\ln(1/t)$. The Page value $S_P$ scales linearly with the system size $L_S$ as shown in the inset of Fig.~\ref{fig: PC}. This indicates volume law EE growth for the Page value. It is important to recall again that although the system size scaling of the steady state entanglement (for example, bipartite EE) is widely explored in literature \cite{Calabrese_2005,Calabrese_2007,KH2013,HA2017,Calabrese2017,Scopa_2021}, such dependence of the Page value is far less understood. We also find that the Page time $t_P \sim L_S^2$. This is a combined consequence of diffusive $\sqrt{t}$ growth of EE and volume law scaling for the Page value.  In other words, $S(t)\sim D\sqrt{t}$ and $S_P \sim a\,L_S$ along with the fact that $S(t_P)=S_P$ yields $t_P\sim L_S^{2}$. This is confirmed in our numerics, where we find excellent quadratic behaviour.

\begin{figure}
    \centering
    \includegraphics[width=0.49\columnwidth]{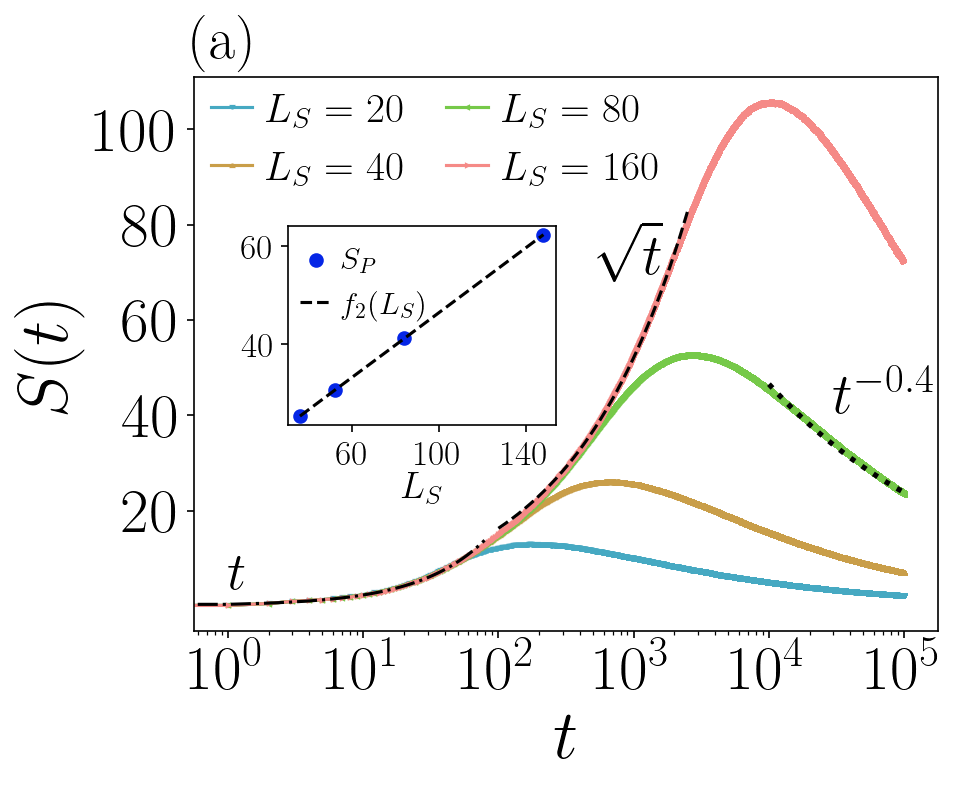}
    \includegraphics[width=0.47\columnwidth]{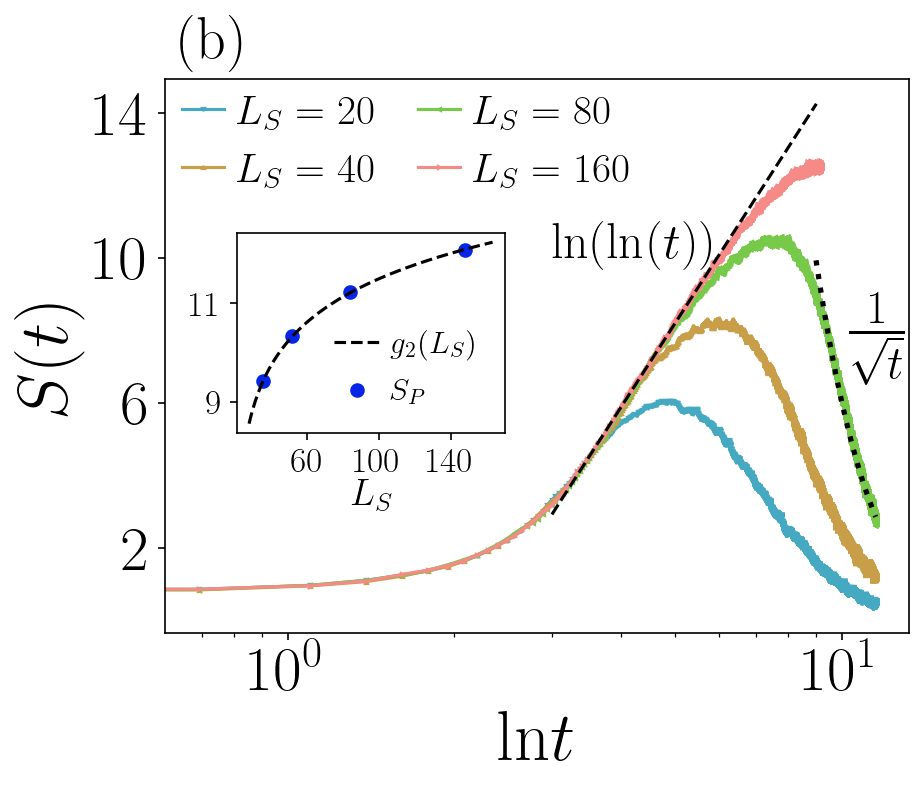}
    \caption{Plot of the entanglement entropy $S(t)$ with time $t$ for the Probe-Probe (PP) setup for the parameters $\gamma =  0.1$, $g=0.5$, $g_c  = 0.4$. The data is averaged over 100 trajectories. (a) SUU case: the entanglement entropy shows a crossover from initial linear growth in time to a diffusive $\sqrt{t}$ growth in the increasing part whereas in the decreasing part, it shows a drop with $t^{-0.4}$, (b) QSD case: the entanglement entropy shows ${\rm ln(ln}t)$ growth in the increasing part followed by a $1/\sqrt{t}$ power law decay in the decreasing part. The inset shows the Page value $S_P$ vs system size $L_{S}$ plot which shows volume law in (a) with $f_{2}(L_S)=a L_S+b$ where $a=0.6$ and $b=1.0$ and sub-volume law in (b) with $g_2({L_S})=a\ln (L_{S})+b$ where $a=3.2$ and $b=-3.4$.}
    \label{fig:PP}
\end{figure}

\begin{figure}
    \centering
    \includegraphics[width=0.49\columnwidth]{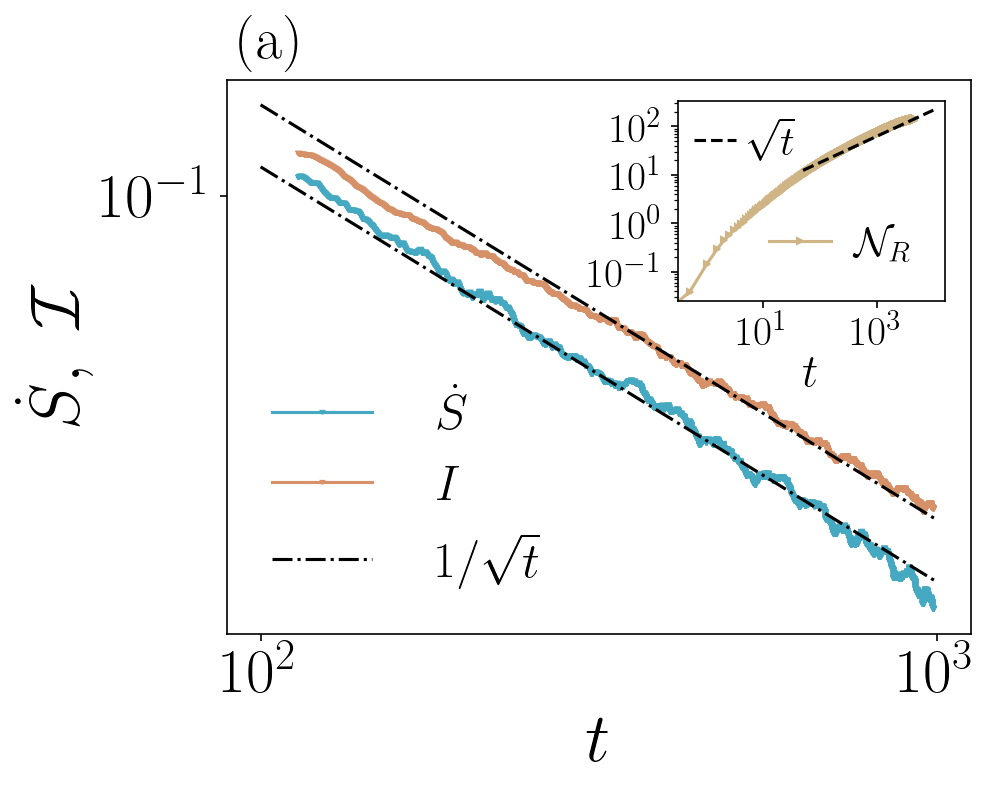}
    \includegraphics[width=0.49\columnwidth]{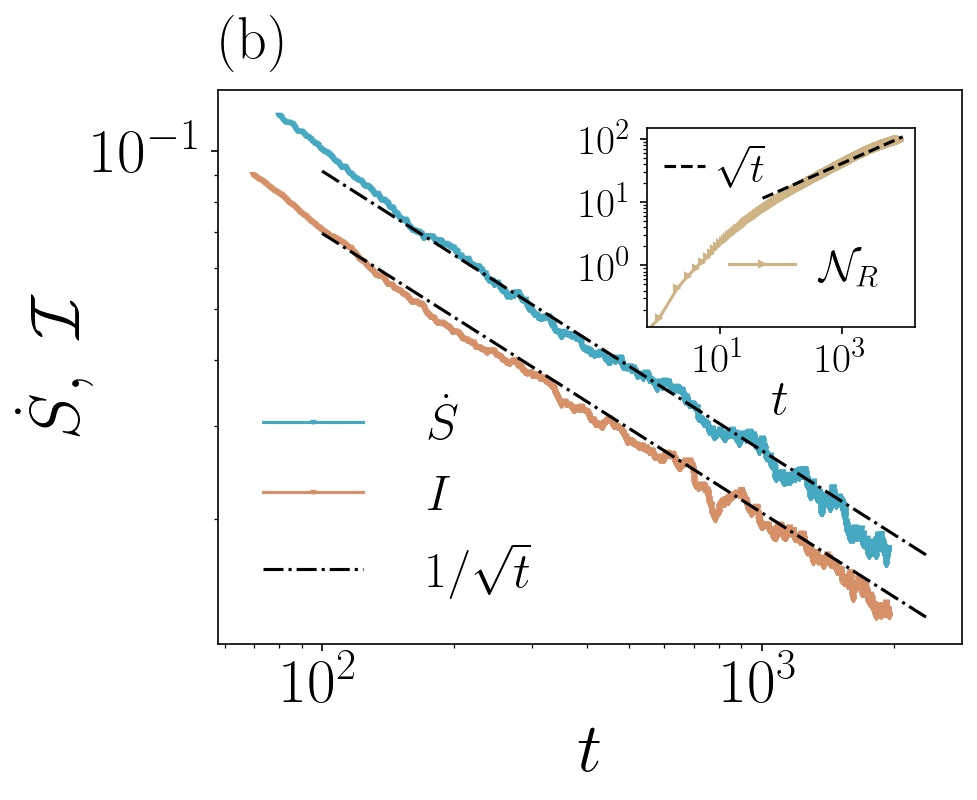}
    \caption{Plot of $\dot{S}(t)= dS(t)/dt$ and $\mathcal{I}(t)= d{\cal N}_R(t)/dt$ with time $t$ for (a) Probe-Clean (PC) case and (b) Probe-Probe (PP) case, under SUU protocol. The plots for both cases show a proportionality relation between $\dot{S}$ and $\mathcal{I}$ till the Page time $t_P$. The inset shows the plot of the number of fermions $\mathcal{N}_R$ in the reservoir with time $t$ which shows the diffusive $\sqrt{t}$ scaling.}
    \label{fig:current}
\end{figure}

We next discuss the Page curve for the PC case under the QSD protocol. This is shown in Fig.~\ref{fig: PC}b. The continuous weak monitoring of the local density results in slower logarithmic growth of EE. This is unlike the SUU protocol, where, due to the absence of physical measurement, faster growth of EE was observed. However, after reaching the Page value $S_P$, the EE decays exactly with the same  $\ln(1/t)$ behaviour, as seen in the SUU case.
Interestingly, the effect of monitoring is clearly manifested in the system-size scaling of the Page value $S_P$. We observe a clear sub-volume $S_P \propto \ln (L_S)$ to area law $S_P \propto L_S^0$ crossover with system size $L_S$ for a fixed monitoring rate $\gamma$, as shown in the inset of Fig.~\ref{fig: PC}b.

The discussions above were for the case when an effectively interacting system was coupled to a non-interacting reservoir. We now discuss the situation where both the system and the reservoir are effectively interacting, i.e., the Probe-Probe (PP) case, schematically shown in Fig.~\ref{fig:Setup_schematic}b. We first discuss Fig.~\ref{fig:PP}a, which corresponds to SUU protocol. Here, we also observe a very early linear growth followed by the diffusive $\sqrt{t}$ behaviour. The EE eventually reaches a maximum Page value ($S_P$) at Page time ($t_P$). We then notice a slower power-law decay, which sharply contrasts the PC case, which showed a faster logarithmic decay. The slower decay in the PP case is rooted in the fact that it takes a much longer time to empty the system due to a large number of scattering effects induced by $L_P=L$ probes. The inset in Fig.~\ref{fig:PP}a shows the volume law behaviour of the Page value, i.e. $S_P \sim L_S$. We further observe the Page time $t_P \sim L_S^2$. In Fig.~\ref{fig:PP}b, we present the PP case with QSD unraveling. Remarkably, in this case, after the very early time linear growth, we find a very slow $\ln(\ln(t))$ growth of EE \cite{PhysRevB.85.094417,PhysRevB.93.205146,SciPostPhys.8.6.083}. The reason for the ultra-slow growth is because of measurement on all the lattice sites, which impedes the growth of EE. After reaching the Page time, the EE decays as a power law similar to the SUU case in Fig.~\ref{fig:PP}a. The inset in Fig.~\ref{fig:PP}b shows a sub-volume system size scaling for the Page value.

It will be interesting to see whether the EE, which involves a careful incorporation of quantum trajectories and is illusive from the point-of-view of the Lindblad master equation, still has some connection with quantities extracted directly from the master equation. This plausible connection is important not only from a theoretical perspective but can also be of significant value for experiments. A proportionality relation between entanglement entropy production rate and particle current~\cite{Kehrein2024, saha2024} has been shown to exist for non-interacting fermionic systems. Remarkably, we observe that under the SUU protocol, for both the PC and the PP cases, 
\begin{equation}
\frac{dS(t)}{dt} \propto {\cal I}(t), \quad t < t_P, \quad {\textrm {for both PC and PP case}}
\label{EE-I_relation}
\end{equation}
as clearly demonstrated in Fig.~\ref{fig:current}(a) and (b), respectively. Eq.~\eqref{EE-I_relation}
holds until the page time. In fact, there is a very early time region ($t\ll t_p$) characterized by linear temporal growth of entropy (i.e., a constant entanglement entropy production rate) which is also consistent with Eq.~\eqref{EE-I_relation}. However, this is not shown in Fig.~\ref{fig:current} since it a small window in the timescale presented. The relation in Eq.~\eqref{EE-I_relation} seizes to exist in monitoring protocols such as QSD, which is somewhat expected since the EE is sensitive to the underlying measurement scheme while the right-hand side of Eq.~\eqref{EE-I_relation} is completely independent of the unraveling procedure.

{\it Summary and Outlook.}-- In this work, we thoroughly examine the entanglement entropy Page curve in an effectively interacting system following the quantum trajectory approach [recall Table.~\ref{table_1}]. Depending on the unraveling protocol and the setup, we observe distinct time dynamics for the growth and decay of EE. The SUU protocol always leads to diffusive growth ($\sqrt{t}$) for EE with a growth coefficient independent of system size. The QSD protocol due to local monitoring always leads to much slower growth of EE. The maximum attainable value of EE, i.e., the Page value, shows interesting system-size scaling (volume law, sub-volume law, area law) depending on the unraveling protocol. The subsequent temporal decay of the Page curve shows distinct behaviour between the two geometrical arrangements of probes. Remarkably, we find strong numerical evidence supporting the relation between the entropy production rate and particle current up to Page time.

It would be interesting to understand the
effect of disorder on the entanglement growth and combined effect of disorder and probes in various geometrical arrangements. Our setup is non-markovian as far as system-reservoir interaction is concerned. However, the dephasing mechanism was assumed to be Markovian. It will be interesting to investigate the effect of non-markovian dephasing mechanisms on the temporal behaviour of entanglement entropy.\\

\noindent 
\textit{Note Added:}
KG, PG, and AN contributed equally to this work.\\

\noindent
{Acknowledgements.}-- M.K. acknowledges support from the Department of Atomic Energy, Government of India, under project No.~RTI4001. B.K.A acknowledges CRG Grant No.~CRG/2023/003377 from the Science and Engineering Research Board (SERB), Government of India. B.K.A. would like to acknowledge funding from the National Mission on Interdisciplinary  Cyber-Physical  Systems (NM-ICPS)  of the Department of Science and Technology (DST), Govt.~of  India through the I-HUB  Quantum  Technology  Foundation, Pune, India. K.G. would like to acknowledge the Prime Minister's Research Fellowship (ID- 0703043), Government of India for funding. K.G. also acknowledges the National Supercomputing Mission (NSM) for providing computing resources of ‘PARAM Brahma’ at IISER Pune, which is implemented by C-DAC and supported by the Ministry of Electronics and Information Technology (MeitY) and DST, Government of India.  B.K.A thanks the hospitality of the International Centre of Theoretical Sciences (ICTS), Bangalore, India under the associateship program.

\bibliography{references.bib}

\onecolumngrid

\setcounter{equation}{0}
\setcounter{figure}{0}
\renewcommand{\theequation}{S\arabic{equation}}
\renewcommand{\thefigure}{S\arabic{figure}}

\newpage

\begin{center}
{\textbf{\underline{Supplementary Material}}}
\end{center}

\section{Details of the unraveling procedures}
\label{sec:ur}
This section is devoted to discussing different unraveling schemes for the dephasing Lindblad equation which we recall to be
\begin{align}
    \frac{d\rho}{dt} = -i[\hat{H}, \rho] + \gamma\sum_{i=1}^{L_P} \left( \hat{n}_i \rho\, \hat{n}_i - \frac{1}{2}\{\hat{n}_i, \rho\} \right),
\label{eq:lindblad_supp}
\end{align}
where $\hat{H}$ is the total Hamiltonian of the setup that includes the system (with size $L_S$) and the reservoir (with size $L_R$) with the total number of sites $L=L_S+L_R$ and it is given by,
\begin{equation}
\label{eq:h_supp}
    \hat{H} = \sum_{i,j =1}^{L-1} h_{ij} \hat{c}^\dagger_{i} \hat{c}_{j}.
\end{equation}
Here, $g$ is the nearest neighbor hopping strength within both the system and the reservoir. However at the junction of the system and the reservoir
we assign a different hopping strength $g_c$. $L_P$ is the total number of probes i.e.,
\begin{align}
L_P = 
\begin{cases}
    L_S \,\,\,\,\,\,\, \text{Probe-Clean (PC)}\,\,{\rm case} \\
    L \,\,\,\,\,\quad \text{Probe-Probe (PP)}\,\,{\rm case,}
\end{cases}
\end{align}
and  $\gamma $ is the dephasing strength. It is to be noted that $\rho$ is the full density matrix of the system and the reservoir.
The dephasing Lindblad equation with the jump operators being the local particle number $\hat{n}_i$ given in Eq.~\eqref{eq:lindblad_supp}, describes the time evolution of the mixed state $\rho(t)$, which is an ensemble average of pure states that evolves stochastically in time, as will be elaborated later.
 Although this noise averaged description is well suited for calculating the expectation values of observables that depend linearly on $\rho(t)$ i.e., $\la \hat{\mathcal{O}} \ra_t =\tr{[\rho(t) \, \hat{\mathcal{O}]}}$, those properties that are inherently nonlinear in $\rho(t)$ fall outside the realm of such averaged description, as in Eq.~\eqref{eq:lindblad_supp}. Notable examples include the  von Neumann entropy $S(t)=-\tr[\rho(t) \ln \rho(t)]$, purity $P=\tr[\rho^2(t)]$, etc. Computation of such quantities, therefore, completely depends on the underlying stochastic dynamics i.e., the unraveling protocol of the Lindblad equation in Eq.~\eqref{eq:lindblad_supp}~\cite{eissler2024}. 
  Different unraveling protocols may result in different values for the same nonlinear functional of $\rho(t)$. Thus, to study such nonlinear functional like entanglement entropy, which is the main focus of this work, it is paramount to discuss the details of the underlying protocol. Given that entanglement entropy is a valid and reliable measure of entanglement only for pure states, it is important to have a description of the mixed-state evolution in terms of pure states. 
  Theoretical descriptions of such pure state dynamics usually appear in terms of stochastic differential equations~\cite{gisin1992quantum,Wiseman1993}, solutions of which are known as quantum trajectories. In this work, we primarily focus on two different kinds of unraveling protocol, namely (i) Stochastic Unitary Unraveling (SUU), (ii) Quantum State Diffusion (QSD) discussed in Subsection.~\ref{subsec:suu} and \ref{subsec:qsd}, respectively.
\subsection{Stochastic Unitary Unraveling (SUU)}
\label{subsec:suu}
\begin{figure}[h!]
    \centering
    \includegraphics[width=1.0\columnwidth]{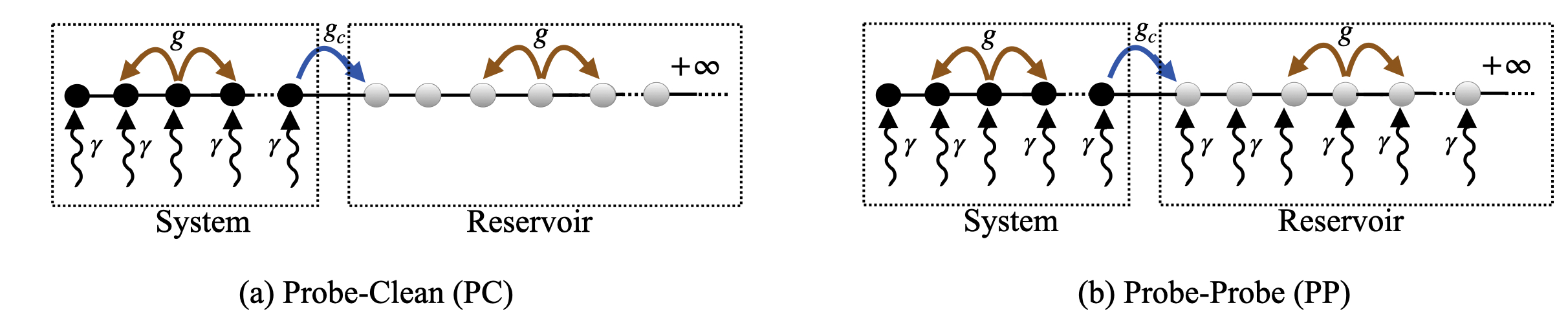}
    \caption{Schematic for the Stochastic Unitary Unraveling (SUU) protocol. We consider a finite-size filled fermionic system of size $L_S$ coupled with an infinite empty fermionic reservoir subjected to onsite fluctuating noise at each site. We call the setup (a) as Probe-Clean (PC) setup, where only the system sites are subjected to noise, with strength $\gamma$ and the reservoir sites are clean, and (b) as the Probe-Probe (PP) setup, where noise is subjected throughout the system and the reservoir lattice sites. The hopping strength for both the system and the reservoir is $g$, except at the junction of the system and the reservoir, where it is $g_c$.  }
    \label{fig:suu_setup_supp}
\end{figure}
 In this subsection, we discuss the Stochastic Unitary Unraveling (SUU)
of the dephasing Lindblad equation in Eq.~\eqref{eq:lindblad_supp}~\cite{PhysRevB.102.100301}. Under this protocol, the lattice system is subjected to onsite fluctuating Gaussian noise $\xi_i (t)$ at each lattice site $i$ [see schematic Fig.~\ref{fig:suu_setup_supp}a and \ref{fig:suu_setup_supp}b for PC and PP setup, respectively]. The stochastic Hamiltonian is therefore given as
\begin{align}
    \hat{H}_T(t) = \hat{H} + \sum_{i=1}^{L_P}\xi_i(t)\,\hat{n}_i, \label{suu-Ham_supp}
\end{align}
where $\hat{H}$ is the Hamiltonian of the full system of size $L$, $\hat{n}_i$ is the local number operator. Recall that $L_P$ is the total number of probes which is $L_S$ in the PC case and $L$ in the PP case. $\xi_i(t)$ in Eq.~\eqref{suu-Ham_supp} is a Gaussian white noise with noise-noise correlation given by $\overline{\xi_i(t)\,\xi_j(t')}=\gamma\,\delta_{ij}\,\delta(t-t')$ with $\gamma$ being the dephasing noise strength. Starting with an arbitrary initial state, the time evolution of the system under the stochastic Hamiltonian $\hat{H}_T(t)$ in Eq.~\eqref{suu-Ham_supp} can be written as,
\begin{align}
    |\psi_{\xi}(t+dt)\ra \approx e^{-i\hat{H}dt-i\sum_{j=1}^{L_P}d\xi^t_j\,\hat{n}_j} |\psi_{\xi}(t)\ra, \label{suu-state_supp}
\end{align}
where $d\xi^t_{j}$ is the infinitesimal noise increment from time $t$ to $t+dt$ and it follows the It\'o rule i.e., $d\xi^t_i\,d\xi_j^t=\gamma\, dt\,\delta_{ij}$ and $\overline{ d\xi_i^t}=0$. The state $|\psi_{\xi}(t)\ra$ for a particular noise realization is called a quantum trajectory. Expanding the exponential in Eq.~\eqref{suu-state_supp} and keeping terms up to $O(dt)$, we obtain,
\begin{align}
    &\big|\psi_{\xi}(t+dt)\big\rangle = \big[\,\hat{\mathbb{I}}-i\hat{H}dt-i\sum_{j=1}^{L_P}d\xi_j^t\,\hat{n}_j-\frac{\gamma}{2}\sum_{j=1}^{L_P}\,\hat{n}_j\,dt\,\big]\,\big|\psi_{\xi}(t)\big \rangle,\label{SSE-SUU_supp}
\end{align}
which can be recast as,
\begin{align}
    &d\,\big|\psi_{\xi}(t)\big\rangle = \big[-i\hat{H}dt-i\sum_{j=1}^{L_P}d\xi_j^t\,\hat{n}_j-\frac{\gamma}{2}\sum_{j=1}^{L_P}\,\hat{n}_j\,dt\,\big]\,\big|\psi_{\xi}(t)\big \rangle.\label{SSE-SUU1_supp}
\end{align}
Eq.~\eqref{SSE-SUU1_supp} is the Stochastic Schr$\ddot{{\rm o}}$dinger Equation (SSE) under SUU protocol. For non-interacting systems, where the Hamiltonian $\hat{H}$ is quadratic, starting with a Gaussian initial state, the Gaussian nature of the state will be preserved under the evolution in Eq.~\eqref{SSE-SUU_supp} for each quantum trajectory. The noise averaged dynamics $\overline{\big|\psi_{\xi}(t)\big\rangle\big\langle\psi_{\xi}(t)\big|}=\rho(t)$ represents the Lindblad equation in Eq.~\eqref{eq:lindblad_supp} which is however non-quadratic in the fundamental fermionic operators.

We will now discuss how to simulate quantum trajectories from Eq.~\eqref{SSE-SUU_supp}. It can be observed that any Gaussian state with a fixed number of fermions can be written as
\begin{align}
    \big|\psi\big\rangle=\prod_{l=1}^{N}\Big(\sum_{j=1}^{L} U_{jl}\,\hat{c}_j^\dagger\,\Big)\,\big|0\big\rangle. \label{state_supp}
\end{align}
Here, $\big|0\big\rangle$ corresponds to the vacuum state of the entire lattice, and $U$ is a $L\times N$ matrix with $N$ being the total number of fermions in the system. It satisfies the isometry $U^{\dagger}\,U=\mathbb{I}_N$, which is required for the norm preservation of the state. Each column of the $U$ matrix represents a single-particle state. For example, the $U$ matrix for the domain-wall state, which is the initial state that we consider in our work, is given by
\begin{align}
    U(0) = \begin{bmatrix}
        1 & 0 & 0 & 0 & \dots &0 \\
        0 & 1 & 0 & 0 & \dots &0 \\
        0 & 0 & 1 & 0 & \dots &0 \\
        \dots & \dots & \dots &\dots &\dots &\dots \\
        \vdots & \vdots & \vdots &\vdots&\dots&\vdots\\
        0 & 0 & 0 & 0 & \dots & 1 \\
        \vdots & \vdots & \vdots &\vdots&\dots&\vdots\\
        \vdots & \vdots & \vdots &\vdots&\dots&\vdots\\
        0 & 0 & 0 & 0 & \dots & 0
    \end{bmatrix},
\end{align}
Now, the time evolution by the SSE in Eq.~\eqref{SSE-SUU_supp} can be written as,
\begin{align}
    \big|\psi_{\xi}(t+dt)\big\rangle \approx e^{\hat{K}_t}\big|\psi_{\xi}(t)\big\rangle, \label{state_kt_supp}
\end{align}
where $\hat{K}_t$ up to $O(dt)$ is the following,
\begin{align}
    \hat{K}_t = -i\hat{H}dt-i\sum_{j=1}^{L_P}d\xi_j^t\,\hat{n}_j-\frac{\gamma}{2}\sum_{j=1}^{L_P}\,\hat{n}_j\,dt. \label{K_t_suu_supp}
\end{align}
Note that the state $\big|\psi_{\xi}(t+dt)\big\rangle$ is the state of the full system, and its dimension is $^L C_N$, thereby making it computationally challenging. Note that $\hat{K}_t$ in Eq.~\eqref{K_t_suu_supp} is quadratic in fermionic creation and annihilation operator, and hence, starting with a Gaussian state, the state at time $dt$ follows from Eq.~\eqref{state_supp}  and \eqref{state_kt_supp},
\begin{align}
   & \big|\psi_{\xi}(dt)\big\rangle =e^{\hat{K}_t}\prod_{l=1}^{N}\Big(\sum_{j=1}^{L} U_{jl}\,\hat{c}_j^\dagger\,\Big)\,\big|0\big\rangle. \label{step_supp}
\end{align}
Taking the term $e^{\hat{K}_t}$ inside the product and the summation, we get
\begin{align}
    & \big|\psi_{\xi}(dt)\big\rangle =\prod_{l=1}^{N}\Big(\sum_{j=1}^{L} U_{jl}\,e^{\hat{K}_t}\,\hat{c}_j^\dagger\,e^{-\hat{K}_t}\,e^{\hat{K}_t}\Big)\,\big|0\big\rangle \label{step2_supp}
\end{align}
Using the fact that $\hat{K}_t|0\ra=|0\ra$, we obtain
\begin{align}
    & \big|\psi_{\xi}(dt)\big\rangle =\prod_{l=1}^{N}\Big(\sum_{j=1}^{L} U_{jl}\,e^{\hat{K}_t}\,\hat{c}_j^\dagger\,e^{-\hat{K}_t}\Big)\,\big|0\big\rangle. \label{step_suu_3_supp}
\end{align}
Using the Baker-Campbell-Hausdorff formula, we get
\begin{align}
e^{\hat{K}_t}\,\hat{c}_j^\dagger\,e^{-\hat{K}_t}=\sum_{k,j}[Me^{-ihdt}]_{kj}\,\hat{c}_k^{\dagger}, \label{sol_suu_supp_1}
\end{align}
where $M$ is a $L \times L$ matrix that takes the form
\begin{align}
M={\rm diag}(e^{-id\xi^0_1+\,\frac{\gamma}{2}\,dt},\,\,e^{-id\xi^0_2+\,\frac{\gamma}{2}\,dt},\,\,\dots \,e^{-id\xi^0_{L_P}+\,\frac{\gamma}{2}\,dt},\,\,1,\dots 1,\,\,1,\dots\,1), \label{M_suu_supp}
\end{align}
where we recall that $L_P$ is the number of probes. $d\xi_i^0$s are independent random Gaussian noise with mean zero and variance $\gamma\, dt$ and $h$ is the single particle $L\times L$ matrix for the Hamiltonian of the system.
From Eq.~\eqref{sol_suu_supp_1} and Eq.~\eqref{step_suu_3_supp}, we obtain,
\begin{align}
    &\big|\psi_{\xi}(dt)\big\rangle =\prod_{l=1}^{N}\Big(\sum_{k,j=1}^{L} U_{jl}\,[Me^{-ihdt}]_{kj}\,\hat{c}_k^{\dagger}\Big)\,\big|0\big\rangle.  \label{lastsetp_suu_supp}
\end{align}
Defining $U(dt) = Me^{-ihdt}\,U(0)$, Eq.~\eqref{lastsetp_suu_supp} can be finally written as
\begin{align}
    &\big|\psi_{\xi}(dt)\big\rangle =\prod_{l=1}^{N}\Big(\sum_{k=1}^{L} U_{kl}(dt)\,\hat{c}^{\dagger}_k\Big)\,\big|0\big\rangle. \label{lastsetp1_suu_supp}
\end{align}
 However, as the $M$ matrix given in Eq.~\eqref{M_suu_supp} contain the factor $e^{\gamma dt}$ in few entries, the $U(dt)$ does not satisfy the isometry condition $U^{\dagger}(dt)U(dt)=\mathbb{I}_{N}$. Hence we perform a QR decomposition of the matrix $U(dt)$ and redefine $U(dt) = Q$ as obtained from QR decomposition.
Thus the evolution of the state $|\psi_{\xi}(dt)\ra$ can be obtained by the matrix $U(dt)$ which is of smaller dimension ($L\times N$). These steps can be followed to generate a quantum trajectory under SUU protocol. 
  The evolution of $U$ in a particular trajectory can be summarized as,
\begin{align}
    U(t+dt)=
    \begin{cases}
    {\rm diag}(e^{-id\xi^t_1}\,\,e^{-id\xi^t_2\,\,}\,\,e^{-id\xi^t_3\,\,}\dots e^{-id\xi^t_{L_S}\,\,}\,\,1 \dots\,1 \,\,1)\,e^{-ihdt}\,U(t), \,\,\,\,\quad {\rm (PC} \,\,{\rm case)}\\
    {\rm diag}(e^{-id\xi^t_1}\,\,e^{-id\xi^t_2\,\,}\,\,e^{-id\xi^t_3\,\,}\dots\,\dots\, e^{-id\xi^t_{L}\,\,}\!\!)\,e^{-ihdt}\,U(t). \,\,\,\,\quad\quad\quad\quad {\rm (PP} \,\,{\rm case)}
    \end{cases}\label{U-pppc_suu_supp}
\end{align}
We now derive an important connection between the $U(t)$ matrix and the correlation matrix defined as $C_{ij}=\la c^{\dagger}_i\,c_j\ra_t$. As the number of fermions $N$ is conserved throughout the dynamics, we can write the state in a trajectory at each instant of time as [obtained from Eq.~\eqref{lastsetp1_suu_supp}]
\begin{align}
    &\big|\psi_{\xi}(t)\big\rangle =\prod_{l=1}^{N}\Big(\sum_{k=1}^{L} U_{kl}(t)\,\hat{c}^{\dagger}_k\Big)\,\big|0\big\rangle,
\end{align}
where $U^\dagger(t)U(t)=\mathbb{I}_N$ for any time $t$. Each of the $N$ columns of $U(t)$ represents a single particle state which are orthogonal to each other, and they form an {\it incomplete} basis in the single particle Hilbert space since $N<L$. We can now elevate the dimension of $U(t)$ from $L\times N$ to $L\times L$ and form a unitary matrix by adding the rest $L-N$ number of single particle basis states. We denote the full matrix as $\mathbb{U}(t)$. Now, in terms of $\mathbb{U}(t)$, the state $\big|\psi_{\xi}(t)\big\rangle $ is given as
\begin{align}
    &\big|\psi_{\xi}(t)\big\rangle =\prod_{l=1}^{N}\Big(\sum_{k=1}^{L} \mathbb{U}_{kl}(t)\,\hat{c}^{\dagger}_k\Big)\,\big|0\big\rangle. \label{bigu_supp}
\end{align}
Now we define another complete set of fermionic creation and annihilation operators as,
\begin{align}
    \hat{b}_{k}^{\dagger}=\sum_{j=1}^{L} \mathbb{U}_{jk}(t)\,\hat{c}_j^{\dagger},\,\,\,\,\,\quad \hat{b}_{k}=\sum_{j=1}^{L} \mathbb{U}^{*}_{jk}(t)\,\hat{c}_j.
\end{align}
The operators $\hat{c}_j$ and $\hat{c}_{j}^{\dagger}$ are also related to these new operators $b_{k}$ and ${b_{k}^{\dagger}}$ respectively as,
\begin{align}
    \hat{c}_{j}^{\dagger}=\sum_{k=1}^{L} \mathbb{U}^{*}_{jk}(t)\,\hat{b}_k^{\dagger},\,\,\,\,\,\quad \hat{c}_{j}=\sum_{k=1}^{L} \mathbb{U}_{jk}(t)\,\hat{b}_k.
\end{align}
The state $\big|\psi_{\xi}(t)\big\rangle$ in Eq.~\eqref{bigu_supp} can also be written in terms of the operator $\hat{b}^{\dagger}_{k}$ as,
\begin{align}
    \big|\psi_{\xi}(t)\big\rangle =\hat{b}_1^{\dagger}\,\hat{b}_2^{\dagger}\,\hat{b}_3^{\dagger}\,\dots \,\hat{b}_N^{\dagger}\big|0\big\rangle
\end{align}
Next we obtain the Correlation matrix $C^{\xi}_{ij}(t)=\big\langle\psi_{\xi}(t)\big|\hat{c}^{\dagger}_{i}\hat{c}_j\big|\psi_{\xi}(t)\big\rangle $ as,
\begin{align}
    \big\langle\psi_{\xi}(t)\big|\hat{c}^{\dagger}_{i}\hat{c}_j\big|\psi_{\xi}(t)\big\rangle = \sum_{l,k=1}^{L} \mathbb{U}^{*}_{il}(t)\,\mathbb{U}_{jk}(t)\,\big\langle0\big|\hat{b}_N\hat{b}_{N-1}\dots \,\hat{b}_1(\hat{b}^{\dagger}_l \hat{b}_k) \hat{b}^{\dagger}_1 \hat{b}^{\dagger}_2\dots \hat{b}^{\dagger}_N\big|0\big\rangle.
\end{align}
We observe that none of the terms corresponding to $l,k>N$ will contribute to the summation. The terms with $l\neq k$ will also not contribute to the summation. Hence only contribution comes from the terms $l=k <N$, thereby resulting in the following form of the correlation matrix $C^{\xi}_{ij}(t)$,
\begin{align}
    C_{ij}^{\xi}(t)=\sum_{l=1}^{N} \mathbb{U}^{*}_{il}(t)\,\mathbb{U}_{jl}(t)= \sum_{l=1}^{N} U^{*}_{il}(t)\,U_{jl}(t) = [U(t)U^{\dagger}(t)]_{ji}.
\end{align}
Thus the correlation matrix has an elegant expression in terms of the $U$ matrices defined in Eq.~\eqref{U-pppc_suu_supp}.
\begin{figure}[h!]
    \centering
    \includegraphics[width=1.0\columnwidth]{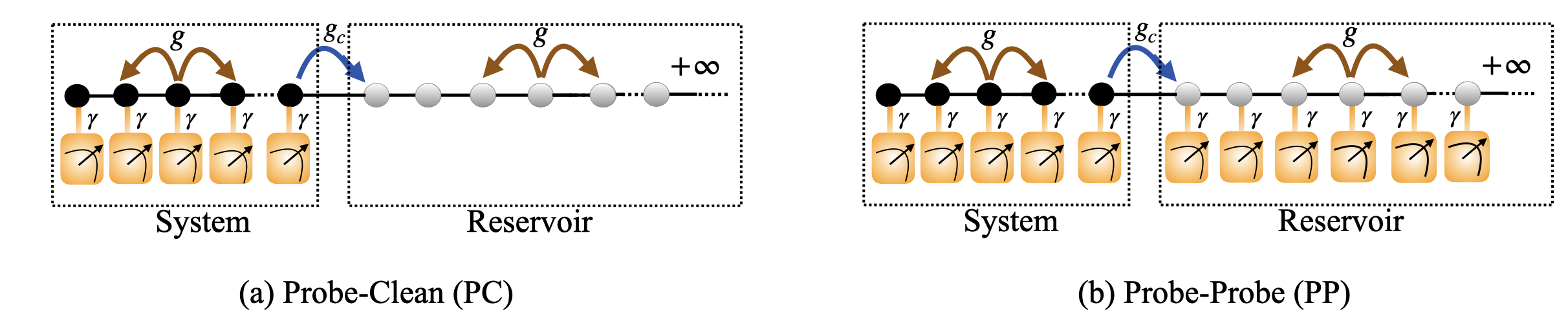}
    \caption{Schematic for the Quantum State Diffusion (QSD) protocol. We consider a finite-size filled fermionic system of size $L_S$ coupled with an infinite empty fermionic reservoir. We call the setup (a) as Probe-Clean (PC) setup, where only the system sites are under continuous weak monitoring of local particle number $\hat{n}_i$ with monitoring rate $\gamma$ and (b) as the Probe-Probe (PP) setup, where both the system and the reservoir sites are monitored. The hopping strength for both the system and the reservoir is $g$, except at the junction of the system and the reservoir, where it is $g_c$.}
    \label{fig:setup_qsd_supp}
\end{figure}
\subsection{Quantum State Diffusion (QSD)}
Another widely employed stochastic protocol that also mimics the same dephasing Lindblad dynamics, given in Eq.~\eqref{eq:lindblad_supp}, upon averaging, is the Quantum State Diffusion (QSD) protocol~\cite{gisin1995,gisin1997,percival1999,Manzano2022,Brun2002,Tamir_qsd,Caves1987,Diosi1998}. Practical implementation of this QSD protocol is possible in ultra-cold atom platforms ~\cite{PhysRevB.110.L081403,Yang2018}. The stochastic dynamics under this protocol describes the evolution of a quantum system subjected to continuous weak measurements~\cite{Quanta14,Jacobs_2006}. Such measurements fall into the class of Positive Operator Valued Measurements (POVM) which are the generalization of projective measurements and it is represented as a set of positive operators $\{\hat{M}_{\mu}\}$ satisfying $\sum_{\mu} \hat{M}^{\dagger}_\mu \hat{M}_{\mu}=\mathbb{\hat{I}}$. Generally, such POVM can be performed by coupling the system with a detector and making projective measurement of the detector. In this subsection, we first derive the $\hat{M}_\mu$ for QSD protocol associated with the Lindblad dynamics in Eq.~\eqref{eq:lindblad_supp}. We next obtain the SSE and further discuss the numerical simulation.

For the specific setup given in Fig.~\ref{fig:setup_qsd_supp}, the system is subjected to continuous weak monitoring of the local particle number $\hat{n}_i$ at $L_P$ number of lattice sites. To derive the SSE under the continuous monitoring, 
we first discretize the time $t$ into $n$ number of intervals of size $dt$, assuming $n$ is very large and $dt$ is very small. Before every measurement, each of the detectors (indexed by $i$), which are used for the purpose of monitoring, is always prepared in a state 
\begin{equation}
    |\phi_i\ra=\int_{-\infty}^{+\infty} dx_i\,\phi(x_i)|x_i\ra,
\end{equation}
 where $x_i$ corresponds to the position of the $i$-th detector. Here $\phi(x_i)$ is the wavefunction of the $i$-th detector which for the case of QSD type monitoring takes the form,
\begin{align}
\phi(x_i)=\Big(\frac{2\gamma\, dt}{\pi}\Big)^{1/4}e^{-\gamma \,dt\,x_i^{2}}, \label{dec_wavefunc_supp}
\end{align}
where $\gamma$ is the monitoring rate which is the same as the dephasing strength in the Lindblad equation in Eq.~\eqref{eq:lindblad_supp}. 

The initial state of the full system is expressed in the local number operator basis and it is given as,
\begin{align}
|\psi(0)\ra=\sum_{n_1,n_2,\dots n_L}F_{n_1\dots n_L}(0) \,|n_1\ra|n_2\ra\dots|n_L\ra, \label{ini_state_supp}
\end{align}
where $|n_i\rangle$ corresponds to the eigenvector of local $\hat{n}_i$ operator and $F_{n_1\dots n_L}$ is,
\begin{align}
F_{n_1\dots n_L}=\la n_1\dots n_L|\psi(0)\ra.
\label{eq:F}
\end{align}
We now write the joint state of the detectors and the system which belongs to the Hilbert space $\mathcal{H}_{\rm SD}=\mathcal{H}_{D_1}\otimes\mathcal{H}_{S_{1}}\otimes\mathcal{H}_{D_2}\otimes\mathcal{H}_{S_{2}}\otimes\dots\otimes\mathcal{H}_{D_{L_P}}\otimes\mathcal{H}_{S_{L_{P}}}\otimes \mathcal{H}_{S_{L_P+1}}\otimes\dots\otimes \mathcal{H}_{S_L}$ where the subscript `${\rm SD}$' corresponds to the full system and all the detectors and $S_i$ and $D_i$ corresponds to $i$-th lattice site and $i$-th detector, respectively. Therefore the joint state takes the form,
\begin{align}
    |\Psi(0)\ra= \sum_{n_1,n_2,\dots n_L}F_{n_1\dots n_L}(0) \Big[(|\phi_1\ra \otimes|n_1 \ra)\otimes(|\phi_2 \ra \otimes |n_2 \ra) \dots \otimes(|\phi_{L_P}\ra\otimes |n_{L_P}\ra)\Big]\otimes |n_{L_P+1}\ra \dots |n_L\ra. \label{state_before_monitoring_supp}
\end{align}
The ket $|n_{L_P+1}\ra \dots |n_L\ra$ outside square bracket indicate that there are no detectors on those sites. Note that we have used $|\psi(0)\ra$ to represent the state in the system Hilbert space and $|\Psi(0)\ra$ to represent the state in the complete system-detector Hilbert space. To perform the continuous monitoring, we couple the system with the detector by an interaction Hamiltonian $H_{\rm int}$ which is given as, 
\begin{align}
H_{\rm int}(t)=\sum_{j=1}^{L_P}f(t)\,\hat{P}_j \otimes \hat{n}_j,
\end{align}
where $\hat{P}_j$ is the momentum operator of the $j$-th detector and $\hat{n}_j$ is the number operator of the $j$-th site. $f(t) $ is the coupling impulse function which satisfy $\int_{0}^{dt} f(\tau) \, d\tau=1$. One such example of this coupling impulse function can be $f(t)=\sum_{n}\delta(t-n\,\tau_c)$~\cite{sumilan2024} where the coupling time $\tau_c$ lies within $dt/2<\tau_c<dt$. 
Then, the total Hamiltonian of the system and the detector is given as 
\begin{align}
    H_T(t)=\mathbb{\hat{H}}_S+f(t)\sum_{j=1}^{L_P} \,\hat{P}_j\otimes\hat{n}_j,
\end{align}
where $\mathbb{\hat{H}}_S$ corresponds to the system Hamiltonian $H$ given in Eq.~\eqref{eq:h_supp} and it is expressed in the joint Hilbert space $\mathcal{H}_{{\rm SD}}$ of the detectors and local sites by suitable use of identity operators. The corresponding unitary operator for evolving the setup from $t=0$ to $t=dt$ is then given by,
\begin{align}
    \mathbb{U}(dt,0)=\mathcal{T}e^{-i\int_0^{dt}d\tau\,H_T(\tau)} \approx  \mathcal{T} e^{-i\int_0^{dt} f(\tau)\, d\tau \sum_{j=1}^{L_P}\hat{P}_j\otimes\, \hat{n}_j}\, \,e^{-i\,\mathbb{\hat{H}_S}dt}= \Big(e^{-i\hat{P}_1\otimes \hat{n}_1}\otimes\dots \otimes e^{-i\hat{P}_{L_P}\otimes\, \hat{n}_{L_P}} \Big) e^{-i\,\mathbb{\hat{H}_S}dt} \label{unitary_supp}
\end{align}
Here we have used the fact that $\int_{0}^{dt} f(\tau) d\tau=1$ and Eq.~\eqref{unitary_supp} is correct upto $O(dt)$. We first operate the unitary operator $e^{-i\mathbb{H}_Sdt} $ on the state $|\Psi(0)\ra$ given in Eq.~\eqref{state_before_monitoring_supp} and as it only has the system Hamiltonian part it will only change the  $F_{n_1\dots n_L}$ to $\Tilde{F}_{n_1\dots n_L}$. Now operating the unitary operator in Eq.~\eqref{unitary_supp} on the joint state of the detector and the system in Eq.~\eqref{state_before_monitoring_supp}, we obtain,
\begin{align}
    |\Psi(dt)\ra&=\sum_{n_1,\dots,n_L}\!\!\!\Tilde{F}_{n_1\dots n_L}(dt)\Big(e^{-i\hat{P}_1\otimes \hat{n}_1}|\phi_1 \ra \otimes|n_1\ra\Big)\otimes\dots \otimes \Big(e^{-i\hat{P}_{L_P}\otimes \hat{n}_{L_P}}|\phi_{L_P} \ra \otimes|n_{L_P}\ra\Big)\otimes |n_{L_P+1}\ra \dots |n_L\ra \label{Psidt_step1}
\end{align}
Using the fact that $e^{-i\hat{P}_j\otimes \hat{n}_j}|\phi_j \ra \otimes|n_j\ra=e^{-i\hat{P}_1n_1}|\phi_j \ra \otimes|n_j\ra$, we write,
\begin{align}                  
   |\Psi(dt)\ra& =\sum_{n_1,\dots,n_L}\!\Tilde{F}_{n_1\dots n_L}(dt)\Big(e^{-i\hat{P}_1n_1}|\phi_1 \ra \otimes|n_1\ra\Big)\otimes\dots \otimes \Big(e^{-i\hat{P}_{L_P}n_{L_P}}|\phi_{L_P} \ra \otimes|n_{L_P}\ra\Big)\otimes |n_{L_P+1}\ra \dots |n_L\ra . \label{step1_supp}
\end{align}
where
\begin{align}
\Tilde{F}_{n_1\dots n_L}(dt)=\la n_1\dots \ n_L|e^{-iH dt}|\psi(0)\ra.
\end{align}
Inserting the identity operator as a completeness relation for each of the detectors, we obtain
\begin{align}
    |\Psi(dt)\ra=\!\!\sum_{n_1,\dots,n_L}\!\!\!\!\!\!\Tilde{F}_{n_1\dots n_L}(dt)&\Big(\int_{-\infty}^{+\infty}dx_1|x_1\ra \la x_1|e^{-i\hat{P}_1n_1}|\phi_1 \ra \otimes|n_1\ra\Big)\otimes\dots \nonumber\\&\dots\otimes \Big(\int_{-\infty}^{+\infty}dx_{L_P}|x_{L_P}\ra \la x_{L_P}|e^{-i\hat{P}_{L_P}n_{L_P}}|\phi_{L_P} \ra \otimes|n_{L_P}\ra\Big)\otimes |n_{L_P+1}\ra \dots |n_L\ra
\end{align}
The monentum operator $\hat{P}_j$ effectively boosts the wave-function of the $j$-th detector to yield,
\begin{align}
    |\Psi(dt)\ra=\!\!\sum_{n_1,\dots,n_L}\!\!\!\!\!\!\Tilde{F}_{n_1\dots n_L}(dt)&\Big(\int_{-\infty}^{+\infty}dx_1\phi(x_1-n_1)|x_1\ra  \otimes|n_1\ra\Big)\otimes\dots\nonumber\\&\dots\otimes \Big(\int_{-\infty}^{+\infty}dx_{L_P}\phi(x_{L_P}-n_{L_P})|x_{L_P}\ra  \otimes|n_{L_P}\ra\Big)\otimes |n_{L_P+1}\ra \dots |n_L\ra \label{psidt_coupled}
\end{align}
Note that the combined system-detector state $|\Psi(dt)\ra$ in Eq.~\eqref{psidt_coupled} is completely entangled because of the presence of integration over the position variable.
 Finally, we perform the projective measurement of the position operator on every detector. Mathematically it is performed by operating the projection operator 
\begin{align}\mathbb{\hat{P}}_{\mu_1\dots\mu_{L_P}}=\big(|\mu_1\ra\la \mu_1|\otimes\mathbb{\hat{I}}_2\big)\otimes\dots \otimes\big(|\mu_{L_P}\ra\la \mu_{L_P}|\otimes\mathbb{\hat{I}}_2\big)
 \end{align} on the state $|\Psi(dt)\ra$ where $\{\mu_1,\mu_2,\dots, \mu_{L_P}\}$ are the readings from the detectors. Thus the joint state of the system and the detector after every detector is measured is given by,
\begin{align}
|\Psi(dt,\{\mu_i\})\ra=\frac{1}{\mathcal{N}}\,\,\mathbb{\hat{P}}_{\mu_1\dots\mu_{L_P}}\big|\Psi(dt)\big\rangle=\frac{1}{\mathcal{N}}\!\!\!\sum_{n_1,\dots,n_L}\!\!\!\!\!\!\Tilde{F}_{n_1\dots n_L}(dt)&\big(\phi(\mu_1-n_1)\dots\phi(\mu_{L_P}-n_{L_P})\big)\Big[|\mu_1\ra  \otimes|n_1\ra\otimes\dots\nonumber\\&\dots \otimes|\mu_{L_P}\ra  \otimes|n_{L_P}\ra\Big]\otimes |n_{L_P+1}\ra \dots |n_L\ra. \label{decoupled_supp}
\end{align}
Here $\mathcal{N}$ corresponds to the normalization factor of the state $|\Psi(dt,\{\mu_i\})\ra$ which is given as,
\begin{align}
\mathcal{N}^{2}=\la \Psi(dt)|\mathbb{\hat{P}}_{\mu_1\dots\mu_{L_P}}|\Psi(dt)\ra=\sum_{n_1,\dots,n_L}\!\!\big|\Tilde{F}_{n_1\dots n_L}(dt)\big|^{2}\big|\phi(\mu_1-n_1)\big|^{2}\dots\big|\phi(\mu_{L_P}-n_{L_P})\big|^{2} \label{normalization}
\end{align}
From Eq.~\eqref{decoupled_supp}, we see that the system and the detector are not entangled. Therefore if we now take the inner product over the detector subspace $\la \mu_1 \dots \mu_{L_P}|\Psi(dt,\{\mu_i\})\ra$, we can obtain the state of the system i.e.,
\begin{align}
    |\psi(dt,\{\mu_i\})\ra = \frac{1}{\mathcal{N}}\sum_{n_1,\dots,n_L}\Tilde{F}_{n_1\dots n_L}(dt)\big(\phi(\mu_1-n_1)\dots \phi(\mu_{L_P}-n_{L_P})\big)|n_1\ra \dots |n_L \ra. \label{state_sys_supp}
\end{align}
Eq.~\eqref{state_sys_supp} is the state at $dt$ after one measurement and it is conditioned with the outcome $\{\mu_1,\dots,\mu_{L_P} \}$.

As mentioned before, such weak measurements on the system which are done by coupling the system with the detector and making projective measurements on the detector refer to a more general class of measurements called Positive Operator Valued Measurement (POVM). The POVM elements are $\hat{E}_\mu = \hat{M}^{\dagger}_\mu \hat{M}_\mu$ with $\mu=\{\mu_i\}$ and these measurement operators $\hat{M}_\mu(dt)$ operates on the initial state of the system $|\psi(0)\ra$ and produces the state at time $dt$ i.e., $|\psi(dt,\{\mu_i\}\ra$. The form of the measurement operator $\hat{M}_\mu(dt)$ is,
\begin{align}
    \hat{M}_\mu(dt)&=\Bigg(\sum_{n_1,\dots ,n_L}\phi(\mu_1-n_1)\dots \phi(\mu_{L_P}-n_{L_P})\,|n_1\dots n_L\ra\la n_1\dots n_{L}|\Bigg)e^{-i\hat{H}dt}\\&=\Bigg(\Big[\hat{M}_{\mu_1}(dt)\otimes \dots \hat{M}_{\mu_{L_P}}(dt)\Big]\otimes\mathbb{\hat{I}}_2\otimes \dots \otimes\mathbb{\hat{I}}_2\Bigg)e^{-i\hat{H}dt}. \label{M_supp}
\end{align}
Note that the presence of $e^{-i\hat{H}dt}$ in $\hat{M}_\mu(dt)$ is due to the fact that within the interval $dt$, the system is also evolving by the Hamiltonian $\hat{H}$.  Here all $\mathbb{\hat{I}}_2$ in Eq.~\eqref{M_supp} which are outside the square bracket correspond to the sites that do not have detectors. Each $\hat{M}_{\mu_j}$ in Eq.~\eqref{M_supp} takes the form,
\begin{align}
    \hat{M}_{\mu_j}(dt)=\sum_{n_j\in \{0,1\}}\phi(\mu_j-n_j)|n_j\ra\la n_j|=c\sum_{n_j \in \{0,1\}}e^{-\gamma \,dt\,(\mu_j-n_j)^{2}}|n_j\ra\la n_j|, \label{mmu_j_supp}
\end{align}
where $c=(2\gamma\, dt/\pi)^{1/4}$. Recall that the wave-function $\phi(x_j)$ is given in Eq.~\eqref{dec_wavefunc_supp}.
Each $\hat{M}_{\mu_j}$ satisfies the relation $\int _{-\infty}^{\infty}d\mu_j\,\hat{M}^{\dagger}_{\mu_j}\hat{M}_{\mu_j}=\mathbb{\hat{I}}_2$. From Eq.~\eqref{mmu_j_supp}, we now write the operator $\hat{M}_\mu(dt)$ as,
\begin{align}
    \hat{M}_{\mu}(dt)=c^{L_P}e^{-\gamma dt\sum_{j=1}^{L_P}(\mu_j-n_j)^{2}}e^{-i\hat{H}dt} \approx c^{L_P}e^{-i\hat{H}dt}e^{-\gamma dt\sum_{j=1}^{L_P}(\mu_j-n_j)^{2}}
\end{align}
where we have used the fact that $e^{-iAdt}\,e^{-i Bdt}\approx e^{-iBdt}\, e^{-iAdt}+ O(dt^{2})$ because we are interested only upto $O(dt)$.
Thus the state at time $dt$ in terms of the operator $\hat{M}_\mu(dt)$ can be approximated as,
\begin{align}
|\psi(dt,\{\mu_i\})\ra=\frac{1}{\mathcal{N}_c}\hat{M}_\mu(dt)|\psi(0)\ra\approx \frac{c^{L_P}}{\mathcal{N}_c}e^{-i\hat{H}dt}e^{-\gamma dt\sum_{j=1}^{L_P}(\mu_j-\hat{n}_j)^{2}}|\psi(0)\ra, \label{psitdt_supp}
\end{align} 
which is correct upto $O(dt)$. Recall that $|\psi(0)\ra$ is the initial state given in Eq.~\eqref{ini_state_supp}. Here
 $\mathcal{N}_c$ is the normalization factor for the state $|\psi(dt,\{\mu_i\})\ra$ in Eq.~\eqref{psitdt_supp} which is given as,
\begin{align}
    \mathcal{N}_c= \la\psi(0)|\hat{M}^{\dagger}_{\mu}(dt)\hat{M}_{\mu}(dt)|\psi(0)\ra,
\end{align}
with $\hat{M}_{\mu}(dt)$ being,
\begin{align}
    \hat{M}_\mu(dt)\approx c^{L_P}\, e^{-i\hat{H}dt}e^{-\gamma dt\sum_{j=1}^{L_P}(\mu_j-\hat{n}_j)^{2}}.
\end{align}
Eq.~\eqref{psitdt_supp} is the conditional state $|\psi(dt,\{\mu_i\})\ra$ at time $dt$ conditioned on the measurement outcomes $\{\mu_j\}$. Next we calculate the probability of obtaining this conditional state $|\psi(dt,\{\mu_i\})\ra$ given in Eq.~\eqref{psitdt_supp},
\begin{align}
    P(\mu_1,\dots,\mu_{L_P};dt)&= \mathcal{N}_c^2=\la\psi(0)|\hat{M}^{\dagger}_{\mu}(dt)\hat{M}_{\mu}(dt)|\psi(0)\ra.
\end{align}
Using $|\psi(dt,\{\mu_i\})\ra$ given in Eq.~\eqref{psitdt_supp}, we get
\begin{align}
    P(\mu_1,\dots,\mu_{L_P};dt)=\sum_{\{n_j\}}|F_{n_1\dots n_L}(0)|^{2}\prod_{j=1}^{L_P}\Big(\frac{2\gamma dt}{\pi}\Big)^{\frac{1}{2}} e^{-2\gamma dt (\mu_j-n_j)^{2}}. \label{jpdf_supp}
\end{align}
Eq.~\eqref{psitdt_supp} and~\eqref{jpdf_supp} are the main ingredients to construct the SSE. Repeating such conditional evolution through the operator $\hat{M}_{\mu}(dt)$ with associated probability $P(\{\mu_i\};t)$ in each interval $dt$, we can obtain the state $|\psi(dt,\{\mu_i\})\ra$ at any time $t$. Note that all these $\mu_j$ are not independent random variables rather they are described by the joint probability distribution function $P(\{\mu_i\};t)$ in Eq.~\eqref{jpdf_supp}. From Eq.~\eqref{jpdf_supp}, we get the distribution of each $\mu_j$ as,
\begin{align}
    p(\mu_j,dt)&=\int_{-\infty}^\infty d\mu_1 \dots\int_{-\infty}^\infty  d\mu_{j-1}\int_{-\infty}^\infty d\mu_{j+1} \dots\int_{-\infty}^\infty d\mu_{L_P} P(\mu_1,\dots,\mu_{L_P};dt) \\
    &=\sum_{\{n_j\}} |F_{n_1 \dots n_L}(0)|^{2}\Big(\frac{2\gamma dt}{\pi}\Big)^{\frac{1}{2}} e^{-2\gamma dt (\mu_j-n_j)^{2}}
    \end{align}

Now using Eq.~\eqref{eq:F} we can re-express $|F_{n_1\dots n_L}(0)|^2$ which gives,
\begin{align}
    p(\mu_j,dt) &= \sum_{\{n_j\}} \la\psi(0)|n_1\dots n_L \ra \la n_1\dots n_L| \psi(0)\ra \Big(\frac{2\gamma dt}{\pi}\Big)^{\frac{1}{2}} e^{-2\gamma dt (\mu_j-n_j)^{2}} \\
    &= \sum_{\{n_j\}} \la\psi(0)|e^{-2\gamma dt (\mu_j-\hat{n}_j)^{2}} |n_1\dots n_L \ra \la n_1\dots n_L| \psi(0)\ra \Big(\frac{2\gamma dt}{\pi}\Big)^{\frac{1}{2}}.
\end{align}
In the second step we moved the exponential inside $\la\psi(0)|n_1\dots n_L \ra$ by converting $n_j$ into an operator $\hat{n}_j$. Next, we use resolution of identity, $\hat{\mathbb{I}} = \sum_{\{n_j\}}|n_1\dots n_L \ra \la n_1\dots n_L|$ to get,

\begin{align}
    p(\mu_j,dt)&=\Big(\frac{2\gamma dt}{\pi}\Big)^{\frac{1}{2}} \la \psi(0)|e^{-2\gamma dt (\mu_j- \hat{n}_j)^{2}}|\psi(0)\ra\\
    &=\Big(\frac{2\gamma dt}{\pi}\Big)^{\frac{1}{2}} e^{-2\gamma dt (\mu_j-\la \hat{n}_j\ra_0)^{2}}, \label{p_muj_supp}
\end{align}
where $\la \hat{n}_j\ra_0=\la\psi(0)|\hat{n}_j|\psi(0)\ra$. As each $\mu_j$ is a Gaussian stochastic variable, we can write it in terms of a random Gaussian noise as the following,
\begin{align}
    \mu_j = \la \hat{n}_j\ra_0+\frac{d\xi^{0}_j}{2\gamma dt}, \label{muj_random_supp}
\end{align}
where all $d\xi_j^{0}$s are now independent random Gaussian noise with zero mean and variance $\gamma \, dt$ and it satisfies the It\'o rule $d\xi^{0}_i\,d\xi^{0}_j=\gamma\, dt\,\delta_{ij}$. 
Thus using Eq.~\eqref{muj_random_supp} in Eq~\eqref{psitdt_supp}, the state $|\psi(t,\{\mu_i\})\ra$ can be written in terms of independent noise $d\xi^{t}_i$  as,
\begin{align}
    |\psi_\xi(t+dt)\ra&=\frac{c^{L_P}}{\mathcal{N}_c}\exp{\Bigg(-i \hat{H} dt-\gamma dt\sum_{j=1}^{L_P}\Big(\hat{n}_j-\la \hat{n}_j\ra_t-\frac{d\xi^{t}_j}{2\gamma dt}\Big)^{2}}\Bigg) |\psi_\xi(t)\ra \label{exp_supp}
\end{align}
The subscript $\xi$ is introduced to represent the role of noise. We recall that the normalization factor,
\begin{align}
    \mathcal{N}_c^{2} = \sum_{\{n_j\}}|F_{n_1\dots n_L}(0)|^{2}\prod_{j=1}^{L_P}\Big(\frac{2\gamma dt}{\pi}\Big)^{\frac{1}{2}} e^{-2\gamma dt (\mu_j-n_j)^{2}}=c^{2L_P}\,e^{-2\gamma dt \sum_{j=1}^{L_P}(\mu_j-\la \hat{n}_j\ra)^{2}}=c^{2L_P}e^{-\frac{L_P}{2}},
\end{align}
where we have used the expression of $\mu_j$ from Eq.~\eqref{muj_random_supp}.
Now expanding the square inside the exponential in Eq.~\eqref{exp_supp} and putting the form of $\mathcal{N}$, we obtain,
\begin{align}
    |\psi_\xi(t+dt)\ra=\frac{c^{L_P}}{c^{L_P}e^{-\frac{L_P}{4}}}\exp\Big(-i \hat{H} dt-\gamma dt\sum_{j=1}^{L_P}(\hat{n}_j-\la \hat{n}_j \ra_t)^{2}+\sum_{j=1}^{L_P}(\hat{n}_j-\la \hat{n}_j \ra_t)d\xi^{t}_j-\frac{L_P}{4}\Big)
\end{align}
Finally, the state at time $t+dt$ takes the form,
\begin{align}
    |\psi_\xi(t+dt)\ra&=\exp\Big(-i \hat{H} dt-\gamma dt\sum_{j=1}^{L_P}(\hat{n}_j-\la \hat{n}_j \ra_t)^{2}+\sum_{j=1}^{L_P}(\hat{n}_j-\la \hat{n}_j \ra_t)d\xi^{t}_j\Big)|\psi_\xi(t)\ra, \label{psi_final_supp}
\end{align}

Expanding the exponential in Eq.~\eqref{psi_final_supp} and keeping upto terms $O(dt)$, we get,
\begin{align}
    |\psi_\xi(t+dt)\ra=\big[\hat{\mathbb{I}} - i\hat{H}dt+\sum_{j=1}^{L_P}d\xi_j^t\,(\hat{n}_j-\la \hat{n}_j\ra_t)-\frac{\gamma}{2}\sum_{j=1}^{L_P}\,(\hat{n}_j-\la \hat{n}_j\ra_t)^2\,dt\,\big]\,\big|\psi_{\xi}(t)\big \rangle. \label{sse-laststep}
\end{align}
Finally, we write the SSE under the QSD protocol from Eq.~\eqref{sse-laststep},
\begin{align}
    d\,\big|\psi_{\xi}(t)\big\rangle = \big[-i\hat{H}dt+\sum_{j=1}^{L_P}d\xi_j^t\,(\hat{n}_j-\la \hat{n}_j\ra_t)-\frac{\gamma}{2}\sum_{j=1}^{L_P}\,(\hat{n}_j-\la \hat{n}_j\ra_t)^2\,dt\,\big]\,\big|\psi_{\xi}(t)\big \rangle.\label{SSE-QSD_supp}
\end{align}
Here recall that $d\xi_i^t$ is the infinitesimal noise increment in time $dt$ which also follows the It\'o rule i.e., $d\xi^t_i\,d\xi_j^t=\gamma dt \delta_{ij}$ and $\overline{ d\xi_i^t}=0$ and $\la \hat{n}_j\ra_t=\la \psi_{\xi}(t)|\,\hat{n}_j\,|\psi_{\xi}(t)\ra$. Note that the QSD protocol in Eq.~\eqref{SSE-QSD_supp} is nonlinear in $|\psi_{\xi}(t)\ra$ due to the presence of $\la \hat{n}_j\ra_t$ in the evolution which is like the back action of measurement on the system. Such a nonlinearity is absent in the SUU protocol in Eq.~\eqref{SSE-SUU_supp} as this protocol does not describe any physical measurement process. Next, we will discuss the numerical simulation of the SSE under QSD protocol given in Eq.~\eqref{SSE-QSD_supp}.
Similar to SUU protocol, the Gaussianity is still preserved throughout the evolution here as well if one starts with Gaussian initial state such as domain wall state ($|1\,\,1\,\,1\,\dots1\,\,0\,\,0\dots0\ra$) which we had considered in our analysis. Similar to subsection~\ref{subsec:suu} we define $\hat{K}_t$ as,
\begin{align}
    \big|\psi_{\xi}(t+dt)\big\rangle \approx e^{\hat{K}_t}\big|\psi_{\xi}(t)\big\rangle, \label{state_kt_qsd_supp}
\end{align}
From Eq.~\eqref{SSE-QSD_supp}, we write the $\hat{K}_t$ operator in this case,
\begin{align}
    \hat{K}_t&=-i\hat{H}dt+\sum_{j=1}^{L_P}d\xi_j^t\,(\hat{n}_j-\la \hat{n}_j\ra_t)-\frac{\gamma}{2}\sum_{j=1}^{L_P}\,(\hat{n}_j-\la \hat{n}_j\ra_t)^2\,dt \\
    &=-i\hat{H}dt-\sum_{j=1}^{L_P}\la \hat{n}_j\ra_t\Big(d\xi^{t}_j\,+\frac{\gamma}{2} \la \hat{n}_j\ra_t\Big)+\sum_{j=1}^{L_P}\hat{n}_j\Big(d\xi^{t}_{j}+\frac{\gamma}{2}(2\la n_j\ra_t-1)\Big). \label{K_t-qsd_supp}
\end{align}
Hence we can write the state at any instant of time as,
\begin{align}
    \big|\psi_{\xi}(t)\big\rangle=\prod_{k=1}^{N}\Big(\sum_{j=1}^L U_{jk}(t)\,c_j^\dagger\,\Big)\,\big|0\big\rangle,
\end{align}
where recall that $N$ is the total number of particles and $L$ is the total number of sites. The $U$ matrix respects the isometry relation, $U^{\dagger}(t)U(t) = \mathbb{I}_N$. Following procedure similar to Eq.~\eqref{step_supp},\eqref{step2_supp} and \eqref{step_suu_3_supp} in subsection~\ref{subsec:suu}, we get
\begin{align}
e^{\hat{K}_t}\,\hat{c}_j^\dagger\,e^{-\hat{K}_t}=\sum_{k,j}[Me^{-ihdt}]_{kj}\,\hat{c}_k^{\dagger}, \label{sol_suu_supp}
\end{align}
where the matrix $M$ now takes the form
\begin{align}
    M = {\rm diag}(e^{d\xi^t_1+\frac{\gamma}{2}\, (2\la n_1\ra_t \,-1)\,dt},\dots e^{d\xi^t_{L_S}+\,\frac{\gamma}{2}\, (2\la n_{L_{S}}\ra_t \,-1)\,dt},\,\,\,1, \dots\,1, \,\,1)
\end{align}
Again following the procedure of Eq.~\eqref{lastsetp_suu_supp} and \eqref{lastsetp1_suu_supp} in subsection~\ref{subsec:suu} we get the evolution of the $U(t)$ in this case as,
\begin{align}
    U(t+dt)= \begin{cases}
    {\rm diag}(e^{d\xi^t_1+\frac{\gamma}{2}\, (2\la n_1\ra_t \,-1)\,dt},\dots e^{d\xi^t_{L_S}+\,\frac{\gamma}{2}\, (2\la n_{L_{S}}\ra_t \,-1)\,dt},\,\,\,1, \dots\,1, \,\,1)\,e^{-ihdt}\,U(t), \,\,\,\,\,\,\,\quad {\rm (PC} \,\,{\rm case)}\\
    {\rm diag}(e^{d\xi^t_1+\frac{\gamma}{2}\, (2\la n_1\ra_t \,-1)\,dt},\,\dots\,\dots\, e^{d\xi^t_{L}+\frac{\gamma}{2}\, (2\la n_{L}\ra_t \,-1)\,dt\,\,}\!\!)\,e^{-ihdt}\,U(t), \,\,\,\,\,\,\,\,\quad\quad\quad\quad {\rm (PP} \,\,{\rm case)}
    \end{cases}
\end{align}
where $d\xi_i^t$s are independent random Gaussian noise with zero mean and variance $\gamma\, dt$ and $h$ is the single particle Hamiltonian of the system. Having obtained $U(t)$ we can extract the correlation matrix $C_{ij}^{\xi}(t)= [U(t)U^{\dagger}(t)]_{ji}$, where the superscript $\xi$ represents each quantum trajectory. 
\label{subsec:qsd}

\section{Dynamics of bipartite entanglement}
\label{sec:qd}

\begin{figure}[h!]
    \centering
    \includegraphics[width=0.4\linewidth]{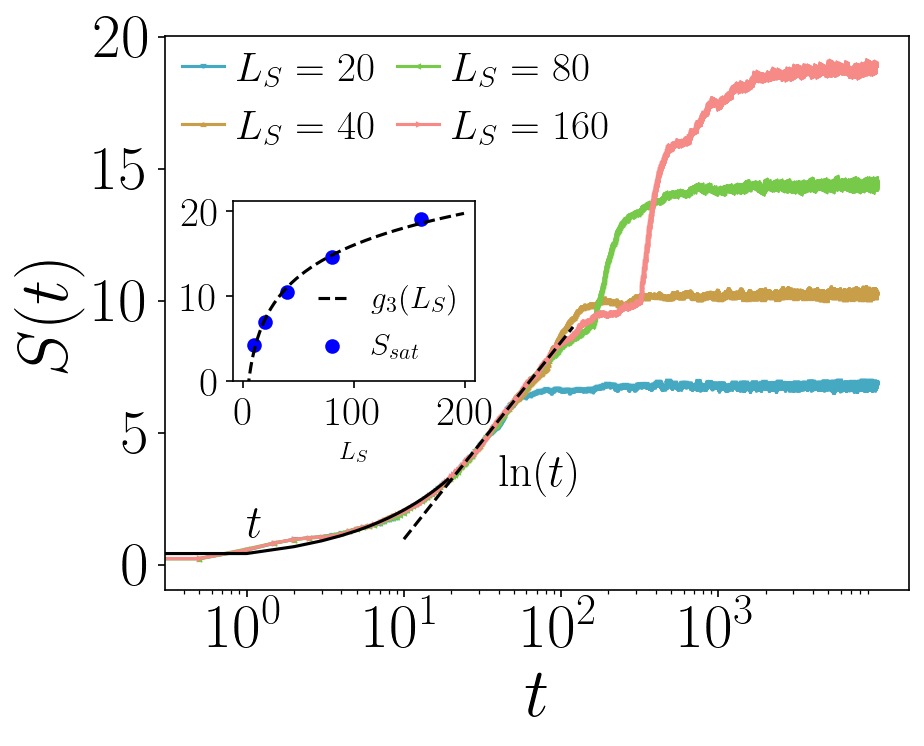}
    \caption{Plot of entanglement entropy $S(t)$ with time $t$ for the bipartite setup for PC case with parameters $\gamma = 0.1$, $g = 0.5$, $g_c = 0.4$. The data is averaged over 100 trajectories. The inset shows the saturation value $S_{\rm sat}$ vs the system size $L_S$. $g_3(L_S) = a \ln(L_S) + b$ with $a = 11.8$  and $b = -8.02$}
    \label{fig:QSD_SS_supp}
\end{figure}

In this section, we discuss the details of EE dynamics for bipartite setup. In such a scenario, instead of a full Page curve dynamics, one sees a growth in EE followed by a steady state saturation. It is important to note that this saturation value may not always match the Page value ($S_P$, the maximum value of EE in a Page curve). It is therefore interesting to understand the saturation dynamics of EE in a bipartite setup. Furthermore, it is important to analyze whether or not the Page value and the saturation value ($S_{\rm sat}$) show similar system size scaling behaviour. In Fig.~\ref{fig:QSD_SS_supp}, we present a scenario where the Page value and the saturation value of EE completely differ and more interestingly show very different system size scaling dependence. We consider the bipartite PC setup with domain wall initial condition and perform QSD unraveling. We set $\gamma = 0.1$. Initially, the EE shows similar behaviour to the EE in the large reservoir case (see Fig.~\ref{fig: PC}b) i.e. starting from zero it shows a logarithmic growth until it reaches a maximum value. But contrary to the large reservoir case, in the bipartite case, the finite size effect kicks in. The particle front that flows ballistically in the empty reservoir, reflects off the boundary (due to the bipartite nature of the setup) and flows back to the system. Around this time scale there is also a sharp increase in the EE value as seen in Fig~\ref{fig:QSD_SS_supp}. It is followed by the saturation of EE value. It is clear that this saturation value is completely different than the Page value (see Fig.~\ref{fig: PC}b). Moreover, as seen in the inset of Fig~\ref{fig:QSD_SS_supp},  the EE in this bipartite setup clearly shows a sub-volume law scaling up to $L_S = 160$ for $\gamma = 0.1$. In contrast,  in Fig.~\ref{fig: PC}b inset, the Page value for system-reservoir setup shows a sub-volume to area law crossover with crossover seen at $L_S \approx 160$ for the same $\gamma=0.1$ value.

\section{Initial condition dependence on entanglement entropy}
\label{sec:ic}

\begin{figure}[h!]
    \centering
    \includegraphics[width=0.4\linewidth]{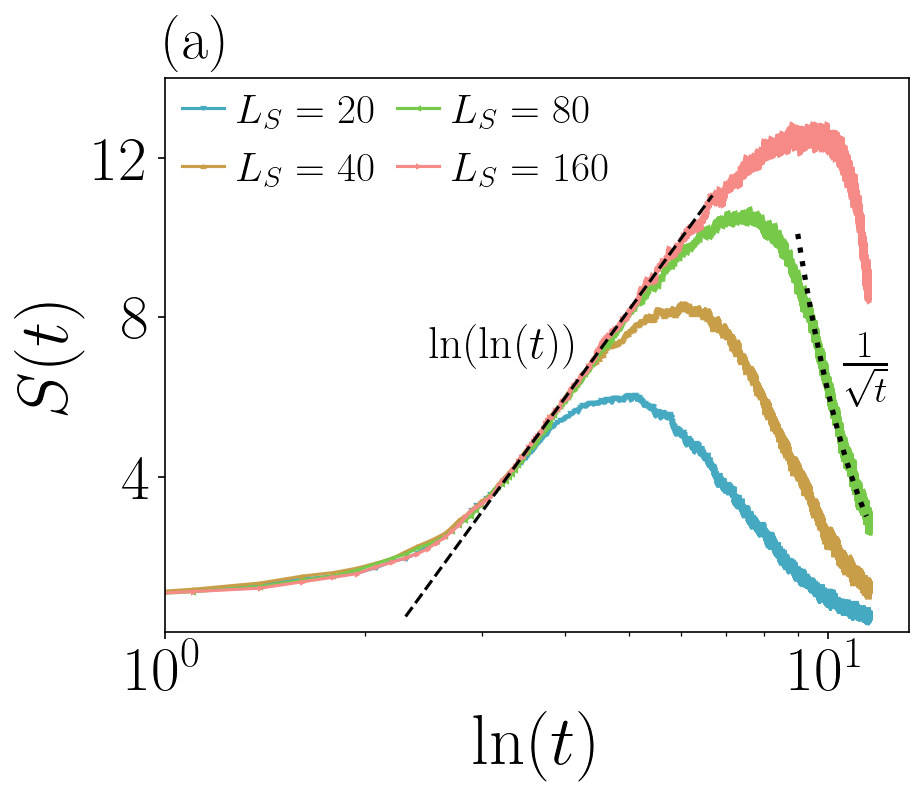}
    \includegraphics[width=0.4\linewidth]{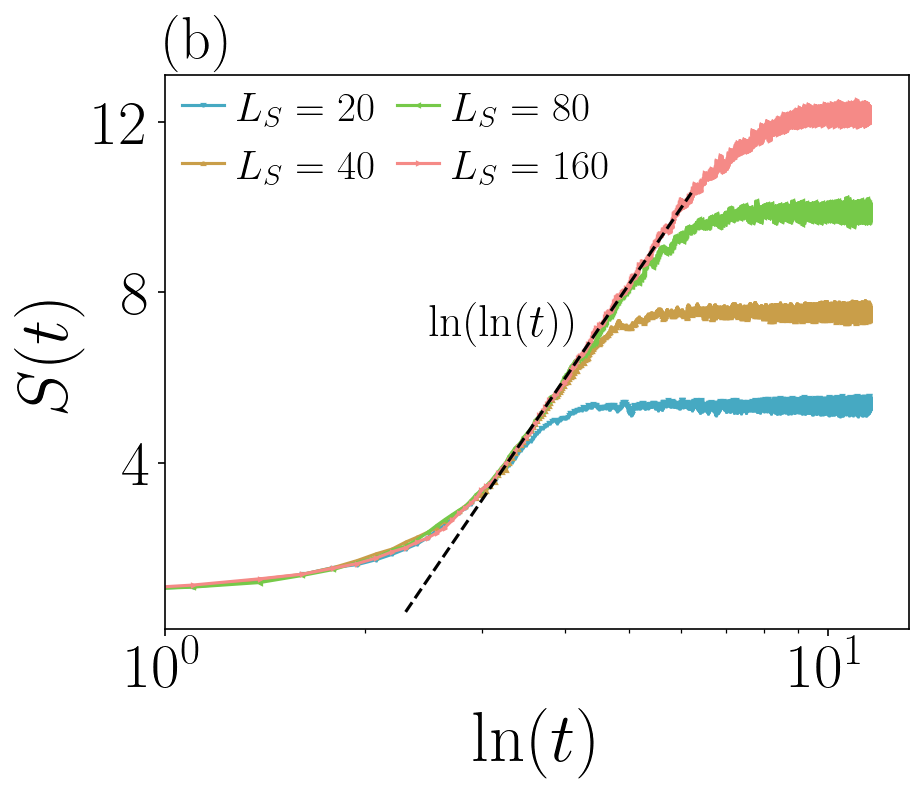}
    \caption{Plot of entanglement entropy $S(t)$ with time $t$ for the Neel state initial condition for the Probe-Probe (PP) setup with parameters $\gamma = 0.1$, $g = 0.5$ and $g_c = 0.4$. (a) The setup considered in the main text i.e., the case when the reservoir is much larger than the system. (b) The setup with bipartite configuration i.e., the system and its complement are of equal size.}
    \label{fig:Neel_QSD_supp}
\end{figure}

The question of how our results of entanglement entropy depend on the details of the initial condition is an interesting question that requires a thorough investigation. In this direction, we use another initial condition which is a Neel state. More precisely, in our setup by Neel state we mean the following: the system sites are filled alternatingly whereas the reservoir is completely vacuum.  In this section,  we discuss the dynamics of EE in the PP case with QSD unraveling starting with a Neel state and with $\gamma = 0.1$. As seen in Fig~\ref{fig:Neel_QSD_supp}a, the Neel state initial condition also demonstrates Page curve dynamics. The rising part of EE shows a $\ln(\ln(t))$ growth. Subsequently, the EE reaches the Page value and decays as a power-law. On comparing the PP case with the domain wall initial condition, as discussed in the main text (see Fig~\ref{fig:PP}b), the Neel state initial condition shows the same growth and decay in time for EE. In Fig~\ref{fig:Neel_QSD_supp}b we plot the bipartite EE which also shows the $\ln(\ln(t))$ growth and finally saturates to a system-size dependent value. 


\end{document}